\documentclass[letterpaper,twocolumn,10pt]{article}
\usepackage[available,functional,reproduced]{usenixbadges}
\usepackage{usenix}
\usepackage{mystyle}

\usepackage{tikz}
\usepackage{amsmath}

\usepackage{filecontents}

\usepackage{soul}

\begin{document}

\date{}

\title{\Large \bf Semantics Over Syntax: Uncovering Pre-Authentication 5G Baseband Vulnerabilities}







\author{
{\rm Qiqing Huang}\\
University at Buffalo
\and
{\rm Xingyu Wang}\\
University at Buffalo
\and
{\rm Wanda Guo}\\
Texas A\&M University
\and
{\rm Guofei Gu}\\
Texas A\&M University
\and
{\rm Hongxin Hu}\\
University at Buffalo
} 

\maketitle

\begin{abstract}
Modern 5G user equipment (UE) processes Radio Resource Control (RRC) configuration messages during early control-plane exchanges, before authentication and integrity protection are established. Prior work for testing 5G UEs has largely focused on constructing syntactically invalid inputs. In contrast, we show that syntactically valid but semantically inconsistent messages, which violate specification-level field constraints or cross-field dependencies, can drive baseband implementations into invalid states, triggering assertion failures or modem crashes. These findings reveal semantic inconsistencies in pre-authentication signaling as a critical yet underexplored attack surface in 5G UE implementations. 
To address this gap, we present Constraint-Guided Semantic Testing (\ProjectName{}), a framework that systematically extracts specification-level constraints and leverages them to generate targeted semantic violations for testing 5G UEs. \ProjectName{} decodes RRC messages into structured fields, derives schema-based rules, infers cross-field dependencies using a Large Language Model (LLM) in an evidence-bounded manner, and produces syntactically valid test cases that intentionally violate semantic constraints.  
We evaluate \ProjectName{} on both commercial and open-source 5G UEs. On commercial smartphones, it uncovers 7 previously unknown vulnerabilities through responsible disclosure, including 3 high-severity CVEs, affecting 64 chipset models and over 542 commercially available smartphone models. On the open-source OAI UE, \ProjectName{} additionally triggers 29 distinct crash sites.
\end{abstract}
\section{Introduction}

Modern 5G user equipment (UE) processes a sequence of downlink configuration messages during early control-plane exchanges, before authentication and integrity protection are in place. This pre-authentication window allows a nearby adversary with a software-defined radio to deliver crafted messages without credentials, as shown by prior Over-The-Air (OTA) studies~\cite{kim2019touching,park2022doltest,bitsikas2023ue,khandker2024astra,hoang2025llfuzz,rupprecht2016putting}. Prior efforts largely target memory corruption or syntax-level mutations~\cite{garbelini2023_5ghoul,potnuru2021berserker}. Yet a critical blind spot remains: \emph{semantic inconsistencies} in configuration messages. These inconsistencies arise when fields violate specification-level constraints or cross-field relations while still passing ASN.1 parsing, allowing them to reach the baseband implementation. Once exercised, such violations may push the baseband into invalid states, triggering assertion failures or modem crashes and giving rise to denial-of-service conditions.

However, existing approaches are ill-suited to address this emerging attack surface.
Over-the-air testing generates syntactically valid inputs but typically relies on random or grammar-based mutations, which are limited in their ability to exercise the field-to-field dependencies mandated by 3GPP standards~\cite{garbelini2023_5ghoul,park2022doltest,hoang2025llfuzz}. 
Emulation and reverse-engineering frameworks provide valuable visibility into baseband implementations~\cite{hernandez2022firmwire,klischies2025basebridge,ranjbar2025stateful,kim2021basespec,kim2023basecomp,golde2016breaking,grassi2020exploring,grassi2018exploitation}, but they lack a systematic approach for translating specification-level semantic constraints into concrete testing objectives.

To discover these semantic inconsistencies, we need to extract semantic relations among fields in Radio Resource Control (RRC)  messages. 
One challenge is that such relations only become visible after configuration messages are decoded into information elements (IEs) and fields through ASN.1 decoding. 
Even then, the evidence for these relations is \textit{scattered} across multiple 3GPP documents, including schema definitions in TS~38.331~\cite{3gpp38.331} and normative clauses in other specifications~\cite{3gpp38.211,3gpp38.213,3gpp38.214,3gpp38.321,3gpp38.322}, and is expressed primarily in natural language. 
For example, TS~38.331 alone has over a thousand pages, and cross-document references, field aliases, release guards, and conditional optionals introduce further ambiguity. 
This dispersion and informality make the systematic extraction of semantic constraints from normative text difficult. 
While deterministic parsing can gather ASN.1 blocks and normative snippets, deciding whether two fields are semantically dependent requires reasoning that goes beyond fixed rules. 
Large Language Models~(LLMs) are well-suited to this reasoning step: they combine natural-language understanding with emerging inference capabilities~\cite{openai2023gpt4,bubeck2023sparks, patir2025towards}, making them a promising tool for uncovering semantic dependencies in normative text and translating them into structured rules.

In this paper, we present \emph{Constraint-Guided Semantic Testing (\ProjectName{})}, 
a framework that systematically extracts specification-level constraints and uses them to construct semantic violations for testing 5G UE implementations. We define a four-class taxonomy of field-level constraints: (i) field value ranges and (ii) field presence, which can be deterministically derived from TS~38.331 schema definitions~\cite{3gpp38.331}; and (iii) intra-IE field dependencies and (iv) inter-IE field dependencies, which are scattered across multiple 3GPP documents~\cite{3gpp38.211,3gpp38.213,3gpp38.214,3gpp38.321,3gpp38.322} and often expressed implicitly in natural language. The latter two categories are especially challenging because they span hundreds of pages of prose and involve cross-field relations that are not directly machine-readable. 
To address this challenge, \ProjectName{} leverages an LLM to extract cross-field semantic relations that are described only in natural-language parts of the 3GPP standards. For each target field or field pair, the model identifies whether the accompanying specification text implies a dependency, and expresses the result in a compact domain-specific language (DSL) rule~\cite{mernik2005and}. The DSL provides a unified, machine-checkable representation of constraints and serves as the interface to test generation. Guided by these rules, our system produces minimal message edits that remain syntactically valid but deliberately violate semantics, making each test case clearly attributable to a specific constraint.\looseness=-1

The key contributions of this paper are as follows:
\begin{itemize}[itemsep=2pt,topsep=2pt,leftmargin=12pt]
\item \textbf{Semantic Constraint Taxonomy.} We introduce a four-class taxonomy of field-level constraints: field value ranges, field presence, intra-IE dependencies, and inter-IE dependencies. The first two classes are derived directly from ASN.1 schema definitions in TS~38.331, while the latter two capture cross-field relations expressed in normative text across the 38.2xx family. By organizing heterogeneous specification rules into this taxonomy, we provide a structured foundation for systematic testing of semantic correctness in 5G UE implementations.

\item \textbf{Constraint-Guided Semantic Testing.} We operationalize the taxonomy into \ProjectName{}, a testing framework that systematically extracts specification-level constraints from 3GPP standards and generates syntactically valid test cases that deliberately violate these constraints. For schema-derived constraints such as value ranges and field presence, \ProjectName{} extracts rules deterministically from TS~38.331. For cross-field dependencies, it leverages an LLM to reason over normative text and identify semantic relations. All cross-field constraints are normalized into a compact DSL.

\item \textbf{Empirical Evaluation.} We demonstrate the practical impact of \ProjectName{} on both commercial smartphones and open-source UE implementations~\footnote{The artifact is available at \url{https://zenodo.org/records/18424200}.}. On commercial smartphones, \ProjectName{} uncovers \textbf{7} previously unknown vulnerabilities, resulting in \textbf{3} high-severity CVEs, with the remaining issues currently under vendor triage. The confirmed vulnerabilities affect \textbf{64} chipset models and more than \textbf{542} commercially available smartphone models. In addition, on the 2025 OpenAirInterface (OAI) UE (version \texttt{2025.w05}), \ProjectName{} localizes \textbf{29} unique crash sites with reproducible traces, \textbf{4}~of which have already been fixed upstream.
\end{itemize}



\section{Background}
This section provides background on the 5G UE attack surface, the structure of RRC signaling, and prior observations on implementation flaws that motivate our study.

\subsection{5G Pre-Authentication Attack Surface}

\begin{figure}[h]
    \centering    
    \includegraphics[width=1.0\columnwidth]{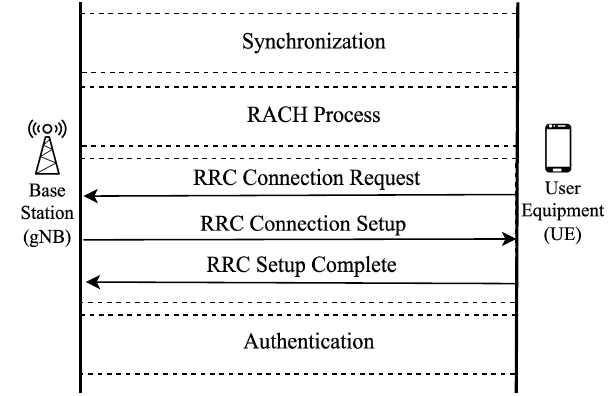}
    \vspace{-5mm}
    \caption{5G Connection Procedure}
    \label{fig:Procedure}
    \vspace{-0.3cm}
\end{figure}

The 5G system includes the radio access network, core network, and user equipment. 
We focus on the UE baseband, which implements the New Radio (NR) stack: PHY/MAC for transmission and scheduling, RLC/PDCP for reliability and ciphering, and RRC for connection setup and configuration, with NAS running above the access stratum to exchange security and session data.
Before mutual authentication (5G-AKA), the UE processes downlink RRC messages in plaintext and without integrity protection (Figure~\ref{fig:Procedure}). 
A rogue gNodeB can thus inject syntactically valid inputs, and prior OTA studies show that such pre-authentication messages can cause crashes or misconfigurations~\cite{park2022doltest,kim2019touching,garbelini2023_5ghoul,hoang2025llfuzz}, motivating our focus on enforcing field-level semantic consistency within RRC.


\subsection{RRC Specification and Semantic Constraints}

5G NR RRC messages are specified in TS~38.331 and encoded using ASN.1 Packed Encoding Rules (PER). Each message consists of nested IEs with typed fields, optional components, and release-specific extensions. While ASN.1 enforces syntactic structure and type domains, many correctness requirements governing UE behavior are defined only through cross-field and context-dependent semantic constraints.

Beyond basic value ranges and optionality enforced by ASN.1, RRC processing depends on a rich set of semantic constraints, including conditional presence rules (e.g., so-called Need codes), intra-IE field dependencies, and cross-IE relationships. These constraints are often described in normative prose across TS~38.331 and the TS~38.2xx family, rather than being enforced by the ASN.1 grammar itself. As a result, a message may be syntactically valid yet semantically inconsistent with the specification, potentially leading to logic-level failures in UE implementations.\looseness=-1

\subsection{Implementation Flaws in 5G Baseband}

Baseband firmware in user equipment must implement complex control-plane and data-plane procedures. Prior work on LTE shows that implementations frequently deviate from the 3GPP standards, producing two recurring categories of flaws~\cite{hussain2021noncompliance,kim2021basespec,chen2023sherlock,park2022doltest}: 
(i) \emph{memory corruption vulnerabilities}, typically from unsafe C/C++ memory handling, and 
(ii) \emph{non-standard-compliant vulnerabilities}, often due to ambiguous specification text or incomplete validation of message syntax. 

\textbf{Memory Corruption Vulnerabilities.}  
Such flaws arise from unsafe pointer arithmetic, insufficient bounds checking, or incorrect allocation, leading to heap/stack overflows and use-after-free. Reported cases include buffer overflows in MediaTek NAS and RRC parsing~\cite{maier2020basesafe}, heap corruption in Samsung when processing malformed \texttt{Mac-MainConfig}~\cite{hussain2021noncompliance}, and integer truncation in 5G NAS updates~\cite{ranjbar2025stateful}. These issues can escalate from DoS to remote code execution.

\textbf{Non-standard-compliant Vulnerabilities.}  
These stem from discrepancies between standard requirements and actual implementations. Studies~\cite{park2022doltest,kim2019touching,kim2021basespec,kim2023basecomp} show that devices may incorrectly process mandatory IEs, accept duplicates, or mishandle invalid headers—leading to authentication bypass, unintended state transitions, or acceptance of plaintext signaling. \looseness=-1

Together, these categories explain most known flaws in LTE and early 5G, largely tied to malformed \textit{syntax} or unsafe memory handling during parsing. 
Regarding recent OTA fuzzers,  5Ghoul~\cite{garbelini2023_5ghoul} applies random RRC mutations without syntax awareness, while LLFuzz~\cite{hoang2025llfuzz} targets LTE lower layers via channel-driven, configuration-aware mutation. 
However, these tools do not explicitly model or target semantic consistency across fields, and therefore are not designed to systematically expose the class of vulnerabilities we study.

In this work, we highlight a distinct class of 5G baseband flaws: \textbf{logic-level semantic inconsistencies}. These arise when an RRC message is syntactically valid but violates specification-level constraints or cross-field relations (e.g., mismatched identifiers, unmet need codes). Such inconsistencies can drive the baseband into invalid internal states in a systematic and reproducible manner, often resulting in assertion failures or modem crashes.

\section{Overview}

\subsection{Scope of Our Work}

We focus on user equipment handling of downlink RRC configuration messages exchanged before integrity protection is activated in 5G Standalone systems. This unauthenticated phase exposes the UE to messages delivered over an untrusted radio link and therefore to spoofed base stations. Our system perturbs such messages while preserving schema compliance and observes UE behavior.

\noindent\textbf{Target Under Test.}
The UE stack is the sole target. We evaluate commercial smartphones via over-the-air delivery, as well as open-source UE implementations under simulation. Our testing focuses on the RRC connection establishment procedure. The standard initial access flow is preserved, and crafted RRC messages are delivered within otherwise compliant RRC control exchanges prior to authentication.

\noindent\textbf{Out of Scope.}
Our analysis does not cover base station or core network implementations, NAS procedures beyond the initial control plane, user-plane traffic, or messages processed under integrity protection. The study focuses on availability-related flaws such as crashes and attach failures, rather than confidentiality or integrity violations.

\subsection{Threat Model}

We consider a practical adversary operating a rogue gNodeB with off-the-shelf SDR hardware (e.g., USRP B210) and open-source software. By emitting stronger downlink signals, the adversary can lure nearby UEs to camp on the fake cell, as shown in prior base station attacks~\cite{park2022doltest,hoang2025llfuzz,garbelini2023_5ghoul}. Without legitimate cryptographic keys, the adversary is limited to the pre-authentication phase, where certain RRC messages remain in plaintext and lack integrity protection. In this window, the attacker can inject malicious control messages, including malformed IEs, reordered or replayed messages, and invalid security headers.

We focus on this threat model as it enables systematic and reproducible testing of unauthenticated RRC message handling. While prior work has explored alternative attacker models capable of direct over-the-air message injection or modification without deploying a rogue base station~\cite{erni2022adaptover, yang2019hiding, ludant2021sigunder, luo2025sni5gect}, the semantic inconsistencies uncovered in our setting are orthogonal to the underlying injection mechanism and are therefore applicable to these stronger adversaries as well.

\begin{figure}[t!]
    \centering    
    \includegraphics[width=1.0\columnwidth]{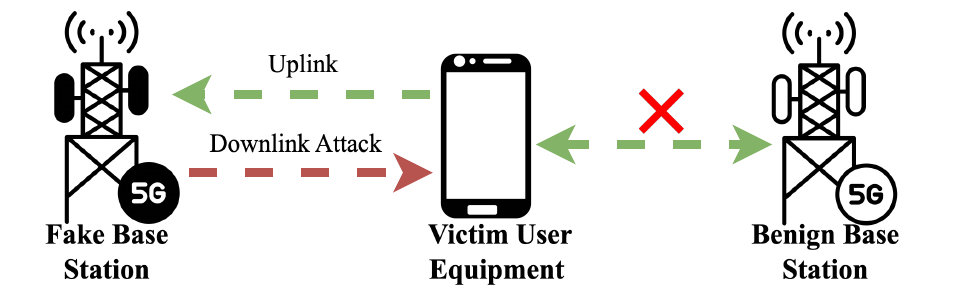}
    \caption{Threat Model}
    \label{fig:threat_model}
\end{figure}

\subsection{Challenges and Approaches}

In this section, we outline the main challenges in building our system and the approaches we adopted to address them, aligned with the design phases.

\subsubsection{Bridging ASN.1 Structure with Semantic Constraints}
5G NR RRC control messages are defined in TS~38.331 and encoded with ASN.1 PER. While existing ASN.1 decoders~\cite{pycrate,asn1c} can parse these messages into structured representations, they expose either low-level bitstreams or high-level object trees, neither of which directly supports constraint-guided mutation. At the bitstream level, unaligned PER's compact encodings and presence bitmaps make blind mutation brittle: a single flipped bit may invalidate the entire message before reaching deeper logic. At the object level, standard decoders lack the annotations necessary to identify which fields participate in which semantic constraints, making it difficult to generate targeted violations.

\noindent\textbf{Our Approach: Dual-Abstraction Representation.}  
We introduce a decoding layer that exposes every message at two complementary levels of abstraction, designed specifically to bridge structure-aware parsing with constraint-guided mutation. The hierarchical IE tree mirrors the ASN.1 grammar and preserves structural context such as optionals and extension markers. The flat table enumerates leaf fields, annotated with canonical paths, current values, presence flags, and ASN.1 domains from TS~38.331. This dual representation enables the mutation engine (§4.3) to precisely target specific fields involved in semantic constraints while maintaining syntactic validity, a capability not provided by standard ASN.1 tools. All later phases, constraint acquisition and mutation, build directly on this structured representation.

\subsubsection{Extracting and Representing Specification Constraints}
Constraints in RRC messages take multiple forms: numeric ranges, presence conditions, and dependencies across IEs. While ranges and presence can be deterministically derived from TS~38.331, intra- and inter-IE dependencies appear only in normative prose across TS~38.2xx documents. These dependencies are described in natural language with cross-references and conditional clauses, making them resistant to deterministic parsing. Without resolving them, many security-relevant rules remain invisible to existing approaches.

\noindent\textbf{Our Approach: Evidence-bound LLM Extraction with DSL Normalization.}  
We employ a large language model~(LLM) in a strictly evidence-bound role to infer such cross-field constraints. For each candidate field pair, the model receives only curated evidence packs that combine ASN.1 fragments with relevant normative sentences. The LLM proposes candidate relations that are then filtered through deterministic gates to discard unsupported inferences. All accepted constraints—whether schema-derived or LLM-inferred—are expressed in a compact domain-specific language (DSL), which unifies heterogeneous rules into a machine-checkable form. This normalization step bridges natural-language specifications with executable test objectives.

\subsubsection{Generating Executable Test Cases}
Having constraints in DSL form does not automatically yield useful tests. A naive violation can make the message unencodable, while editing too many fields obscures attribution. The challenge is to create inputs that remain syntactically valid yet deliberately violate exactly one semantic rule.

\noindent\textbf{Our Approach: DSL-driven Mutation and Faithful Re-encoding.}  
The DSL rules guide the mutation engine by specifying the target fields and conditions to satisfy or violate. 
The engine applies minimal edits within ASN.1 domains so that messages re-encode correctly. 
Constraint types are exercised uniformly: ranges are probed just outside their bounds, presence is tested by toggling optionals, and dependencies are violated through mismatched but valid assignments. 
Each test case remains syntactically valid, tied to one constraint, and produces interpretable outcomes that directly attribute failures to a specific semantic inconsistency.

\begin{figure*}[tb]
    \centering    
    \includegraphics[width=0.95\textwidth]{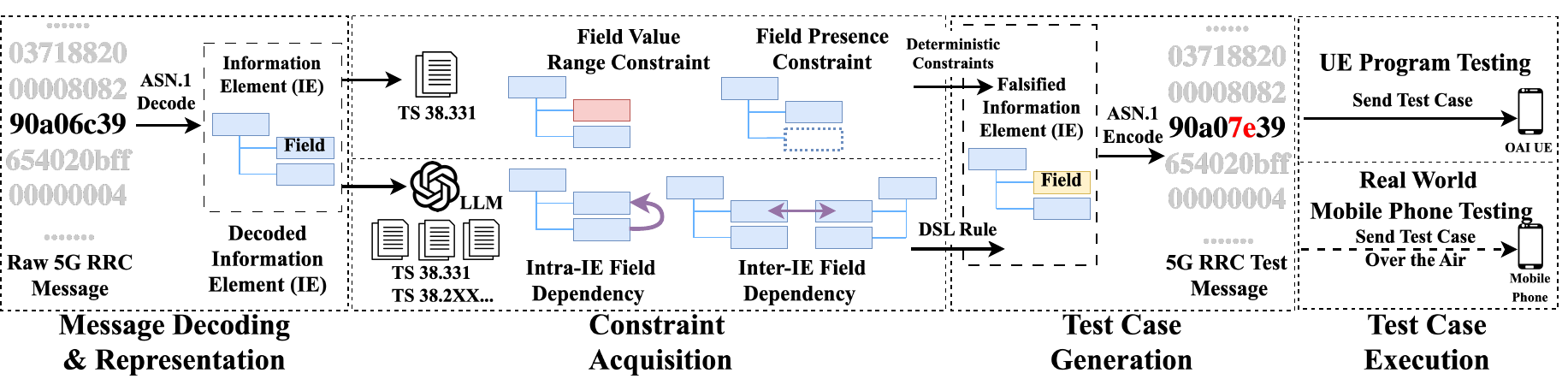}
    \caption{Design Overview}
    \label{fig:overview}
\end{figure*}

\section{Design}


We present a constraint-guided framework for uncovering flaws in 5G RRC implementations. As shown in Figure~\ref{fig:overview}, the system decodes each control message into an IE tree and flat field table, extracts constraints from 3GPP standards, and applies them to synthesize targeted test cases. 



\subsection{Message Decoding and Representation} We decode each RRC message into two complementary representations: a hierarchical IE tree that preserves the ASN.1 structure, and a flat table of leaf fields annotated with their current values, and ASN.1 domains from TS~38.331. This dual representation supports precise, minimal modifications: we leverage an ASN.1 decoding and re-encoding framework to parse RRC messages into field-level representations, enabling fine-grained semantic mutation while preserving schema validity. \looseness=-1

\subsection{Constraint Acquisition}

\subsubsection{Semantic Constraint Taxonomy}

In 3GPP RRC, each message is an ASN.1 structure composed of information elements (IEs). Each IE contains one or more fields with declared domains, and the correctness of message processing depends on respecting a set of semantic constraints. We identify four recurring types:

\noindent\textbf{Field Value Range Constraints.}  
These constraints define permissible numerical intervals or enumerated domains for individual fields. They are typically explicitly stated in specification tables, and violations often lead to out-of-bound access or assertion failures in protocol implementations. Such constraints are straightforward to extract from ASN.1 type definitions or tabulated ranges.

\noindent\textbf{Field Presence Constraints.}  
These constraints specify the required behavior when optional fields are present or absent. In TS~38.331, optional IE fields are annotated with a \emph{need code}, a standardized indicator that prescribes how the UE should handle field absence (e.g., apply a default value, raise an error, or ignore the absence). ASN.1 decoders validate syntactic presence but do not enforce these semantic handling requirements. Incorrect implementation of need code semantics may trigger null pointer dereferences or uninitialized variable access.

\noindent\textbf{Intra-IE Dependency Constraints.}  
These constraints capture semantic dependencies between multiple fields within a single IE, such as index bounds, reference bindings, or logical value relations. They are not enforced by ASN.1 parsers and only surface when implementations check semantics, making them a common source of hidden flaws.

\noindent\textbf{Inter-IE Dependency Constraints.}  
These constraints govern semantic relationships between fields across different IEs or protocol sublayers. They include cross-IE consistency, range alignment, and state dependencies, and typically require reasoning across multiple specifications, which makes them error-prone to implement correctly.


From this taxonomy, field value range and field presence constraints can be derived deterministically from TS~38.331 (5G NR RRC Protocol specification), which defines the RRC ASN.1 schema and value domains~\cite{3gpp38.331}. In contrast, intra-IE and inter-IE field dependencies are specified in normative prose across TS~38.331 and the 38.2xx family, including TS~38.213 (5G NR Physical layer procedures for control)~\cite{3gpp38.213}, TS~38.214 (5G NR Physical layer procedures for data)~\cite{3gpp38.214}, TS~38.321 (5G NR MAC protocol specification)~\cite{3gpp38.321}, TS~38.322 (5G NR RLC protocol specification)~\cite{3gpp38.322} and TS~38.211 (5G NR Physical channels and modulation~\cite{3gpp38.211}). Accordingly, we treat TS~38.331 as the authoritative source for field types and domains, and use the 38.2xx documents as behavioral evidence for extracting cross-field relations.


\subsubsection{Deterministic Constraint Extraction}

\noindent\textbf{Field Value Range Constraints.}
We deterministically derive the declared numeric bounds for each field from TS~38.331 and map every concrete field instance in the decoded message to its corresponding ASN.1 rule. Messages are first decoded into an information-element tree and flattened into leaf instances, so a mutation can target exactly one field with precise scope while leaving the surrounding structure unchanged.



\noindent\textbf{Field Presence Constraints.}
In TS~38.331, optional IE fields are annotated with \emph{need codes} that prescribe the UE's behavior when the field is absent. The two primary need codes are Need~M (maintain), which requires the UE to reuse the last configured value for that field, and Need~R (remove), which requires the UE to clear any previously configured value. We extract these annotations directly from the ASN.1 schema in TS~38.331 and generate test cases that toggle field presence to verify correct need code handling. For instance, testing whether an implementation properly maintains or clears values when expected fields are absent.

\begin{figure}[t!]
    \centering    
    \includegraphics[width=1.0\columnwidth]{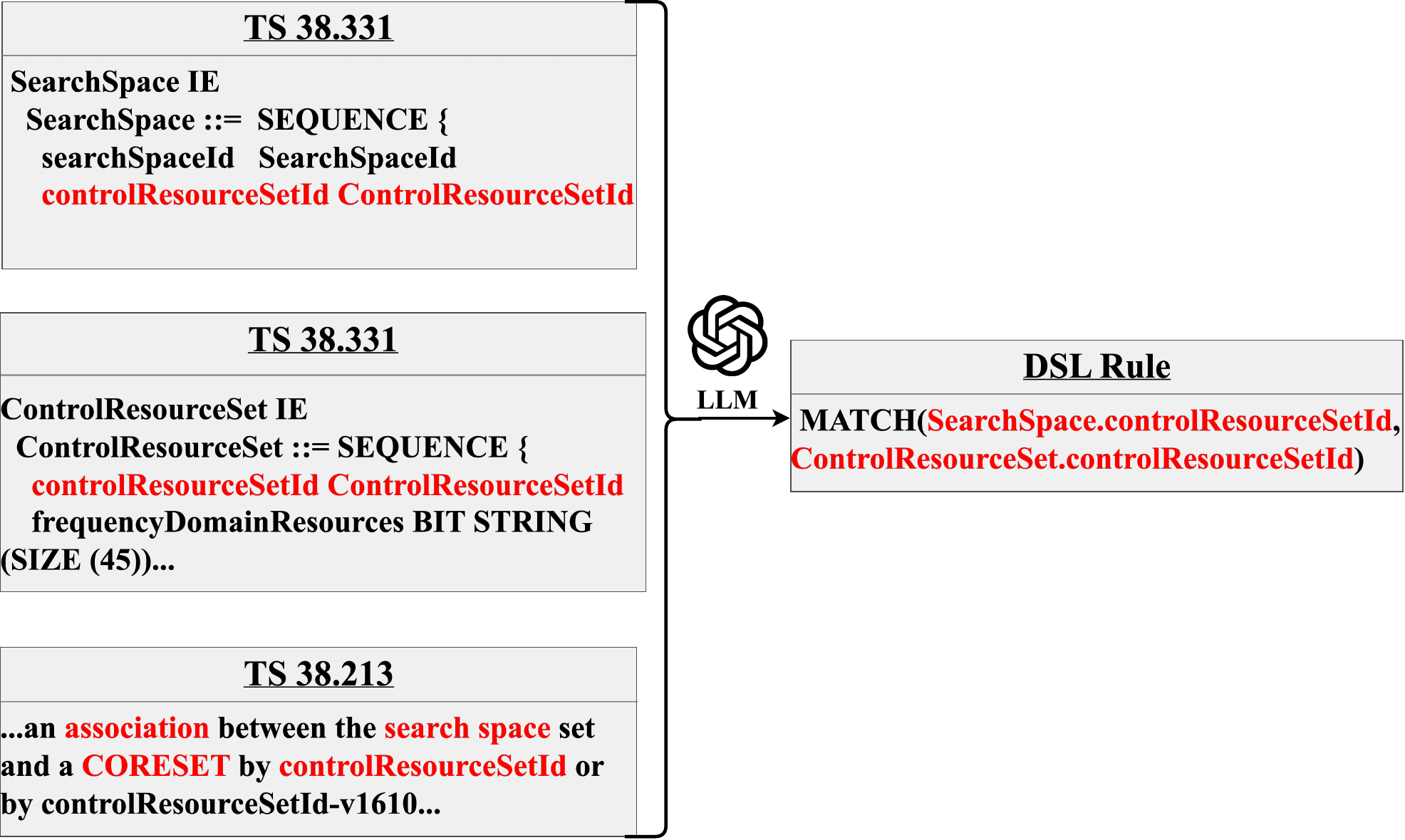}
    \caption{Example of Evidence-Bound DSL Rule Induction}
    \label{fig:DSL_example}
\end{figure}



\subsubsection{LLM-based Dependency Constraint Acquisition}

We extract constraints at two levels. 
\emph{Intra-IE constraints} govern dependencies between fields within the same Information Element. 
\emph{Inter-IE constraints} enforce reference integrity across different IEs.

\noindent\textbf{IE Selection and Field Pair Identification.}
For intra-IE constraints, we apply greedy set cover to select 83 target IEs from 299 candidate IEs, achieving 96.38\% coverage of message fields and yielding 7,019 field pairs for analysis.
For inter-IE constraints, we identify 32 reference fields via pattern matching on names ending with \texttt{Id} or \texttt{ID}, then enumerate 3,178 cross-IE reference relationships where a field must reference an entity defined in another IE with a matching base name. Greedy selection of candidate IEs achieves 94.56\% coverage of these reference pairs. These counts characterize candidate relationships identified during the constraint acquisition stage, prior to evidence packaging and LLM-based rule induction.

\noindent\textbf{Evidence Extraction and Field Mapping.}
We systematically extract normative sentences from 3GPP specifications using keyword-based scanning across multiple categories: mandatory indicators (\emph{`shall'}, \emph{`must'}, \emph{`required'}), conditional markers (\emph{`if'}, \emph{`when'}, \emph{`only if'}), dependency cues (\emph{`depends on'}, \emph{`determined by'}, \emph{`according to'}), and reference patterns (\emph{`correspond'}, \emph{`match'}, \emph{`associate with'}). For each detected keyword, we extract a context window spanning two lines before and after the keyword occurrence to capture complete constraint semantics. Field identification employs multi-pattern matching to handle naming variations: for instance, the ASN.1 field \texttt{schedulingRequestID} may appear in normative text as ``scheduling request ID'', ``scheduling-request-id'', or the abbreviation ``SR''. 
Abbreviations in 3GPP specifications can be overloaded and may refer to multiple related fields or procedures. 
To avoid ambiguous mappings, \ProjectName{} never resolves fields based on abbreviations alone. Abbreviations are treated as weak candidates and are accepted only when the target field pair is uniquely identifiable through co-mention within the same context window and IE scope; otherwise, the evidence is discarded.
We verify field co-occurrence by checking whether both target fields appear within the extracted context window using their respective pattern sets, ensuring that the normative sentence genuinely constrains the field pair. This structured extraction process yields evidence packages with explicit field-to-sentence mappings, which are then forwarded to the LLM for constraint reasoning.

\noindent\textbf{DSL Grammar and Semantics.}
To bridge natural-language specifications and executable test objectives, we express all extracted constraints in a compact, machine-checkable DSL. The DSL is organized along two orthogonal dimensions: \emph{scope} and \emph{semantic family}. At the scope level, constraints are categorized as \emph{intra-IE} (within a single Information Element) or \emph{inter-IE} (across different Information Elements). At the semantic level, all constraints belong to one of two core families. \texttt{ValueDependency} captures conditional relations between fields, including equality, inequality, enumerant mappings, and reference consistency. \texttt{RangeAlignment} expresses coordinated numeric bounds or ordering relations between fields or between a field and a constant. For inter-IE constraints, the DSL provides specialized predicates, such as \texttt{CrossReference}, \texttt{Association}, and \texttt{Conditional}, which are concrete instantiations of \texttt{ValueDependency} across IE boundaries.
Rules are normalized for validation and test generation. Conditional constraints are expressed in implication form \texttt{IMPLIES(<preconditions>, <atom>)}, while unconditional constraints may use atomic predicates for conciseness.

The DSL supports a small, fixed set of canonical operators. 
For value relations, \texttt{ValueDependency} includes equality and inequality (\texttt{EQ}, \texttt{NE}), set membership (\texttt{IN}), enumerant mappings (\texttt{MAP}), simple modular predicates (\texttt{MOD}), and identifier matching (\texttt{MATCH}) as a shorthand for cross-reference consistency checks.

\noindent\textbf{DSL Normalization.}
Normalization takes as input a candidate DSL rule produced by the LLM and applies three deterministic checks to validate and canonicalize it:
(i) \emph{field binding} to resolve referenced placeholders to concrete ASN.1 fields;
(ii) \emph{literal validation} to check admissibility under the corresponding ASN.1 definitions;
and (iii) \emph{canonicalization} to ensure a consistent representation for semantically equivalent rules.
The output of normalization is either the canonicalized DSL rule or \texttt{NO\_RULE} if any mapping is ambiguous or invalid.
Given an intra-IE candidate rule
\texttt{IMPLIES(EQ(field1, `format0'), EQ(field2, 1))}, normalization resolves the placeholders to \texttt{PUCCH-Config.format} and \texttt{PUCCH-Config.nrofPRBs}, validates all referenced values against the ASN.1 definitions, and outputs the canonical rule \texttt{IMPLIES(EQ(format, `format0'), EQ(nrofPRBs, 1))}; otherwise, the rule is rejected as \texttt{NO\_RULE}.

\begin{figure}[b!]
    \centering    
    \includegraphics[width=1.0\columnwidth]{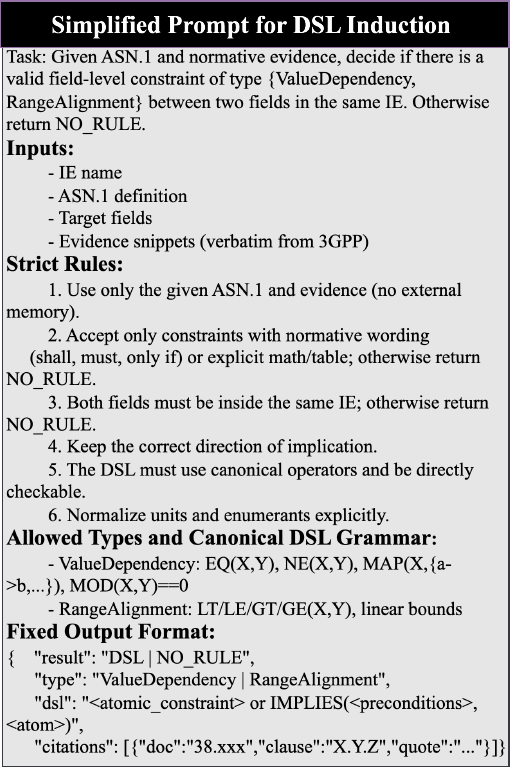}
    \caption{Simplified Prompt for Intra-IE DSL Rule Induction~(Full Version in the Appendix)}
    \label{fig:simplified_prompt}
\end{figure}

\noindent\textbf{Evidence-bound DSL Rule Induction.}
We operationalize this step via a language model in a strictly evidence-bound role. The model reads only the assembled evidence package and must either return a normalized DSL rule or \texttt{NO\_RULE}. Admission is strict: evidence must contain clear normative wording or explicit mathematical/tabular relations. Rules must remain within declared ASN.1 domains, normalize units and enumerants when needed, and include minimal citations. Version or option guards are recorded as preconditions when the text limits scope. Accepted rules are expressed in a compact predicate language with two families: \textit{ValueDependency} (equality/inequality, small enumerant maps, modular relations) and \textit{RangeAlignment} (ordering and linear bounds between fields or between a field and a constant). Items with advisory wording, missing co-mention, tautologies, or empty antecedents are rejected as \texttt{NO\_RULE}.  Figure~\ref{fig:simplified_prompt} illustrates the simplified prompt used for intra-IE constraint induction, where both target fields reside within the same Information Element, and the prompt enforces the same-IE scope. The full prompt template and validation rules are provided in Appendix~\ref{sec:prompt_dsl}. For inter-IE dependency induction, we use the same prompt structure, but allow evidence and field references to span multiple IEs rather than enforcing the same-IE scope.

For example, Figure~\ref{fig:DSL_example} shows an inter-IE field dependency: 
the identifiers \texttt{SearchSpace.controlResourceSetId} and 
\texttt{ControlResourceSet.controlResourceSetId} must be equal. The LLM receives the ASN.1 definitions together with a normative sentence from TS~38.213 and outputs a normalized DSL rule, for example: \texttt{MATCH(SearchSpace.controlResourceSetId, ControlResourceSet.controlResourceSetId)}.



\subsubsection{Rule Application and Minimal Edits}

Given a decoded message and a DSL rule expressed as an implication, the engine plans a minimal change on the named fields to deliberately violate it. Candidate edits are checked against the ASN.1 domains so that the re-encoded message remains well-formed. If no satisfying or violating assignment exists inside the domains, the pair is marked infeasible and skipped. Value range and presence probes are derived deterministically from the schema, while intra- and inter-IE dependency rules use the DSL predicate to guide concrete assignments. Each accepted edit is applied locally, recorded with its field paths and values, re-encoded with \texttt{pycrate}, and delivered through the same execution pipeline. This design ensures that tests remain faithful to the specification and that observed effects can be attributed to one specific violated rule.

\subsection{Test Case Generation}

Given a decoded message, we automatically generate test cases by applying minimal, targeted edits to specific fields according to the extracted DSL constraints, then re-encoding the modified message. Messages that fail to re-encode are discarded. Each test case records its binary payload, edited fields, and the specific constraint it targets for attribution. We execute each test case in both simulation and over-the-air paths under identical monitoring conditions.

\noindent\textbf{Field Value Range.} 
For numeric fields with bounded integer domains, we generate boundary-value tests by setting the field to values just outside the declared bounds. Specifically, one below the minimum and one above the maximum. For example, a field with domain [0, 191] is tested with values -1 and 192. These intentionally violate range constraints to probe robustness, testing whether implementations properly validate input bounds. All other fields remain unchanged.

\noindent\textbf{Field Presence.}
For each optional field annotated with Need~M or Need~R, we create a message variant that \emph{omits} that field (using ASN1VE\cite{asn1ve}) while leaving all other content unchanged, then encode and deliver the message. For Need~M, a crash or null dereference indicates failure to maintain the prior value. For Need~R, a crash that stems from using a cleared value indicates incorrect handling of removal.

\noindent\textbf{IE Field Dependency.}
For dependencies between two fields within one IE or across two IEs, we use an evidence-bound prompt that returns a normalized DSL rule together with a small set of candidate value pairs for the two fields. Before applying any pair, we check types, enumeration membership, and numeric bounds against TS~38.331. When multiple pairs pass these checks, we prefer the one with the smallest deviation from the original values to keep the edit minimal. The selected pair is written to the decoded record and re-encoded as above, yielding a targeted test that keeps the encoding acceptable while intentionally violating the semantic relation.

\noindent\textbf{Seed Selection and Edit Scope.} 
Seed messages are collected during standard RRC procedures, ensuring realistic starting points. For example, consider the inter-IE dependency between 
\texttt{SearchSpace.controlResourceSetId} and 
\texttt{ControlResourceSet.controlResourceSetId}. 
The extracted rule requires the two identifiers to match. From a seed where both fields hold the same value as~2, the generator produces a minimal violating variant by altering \texttt{SearchSpace.controlResourceSetId} to a different valid identifier like 1 while keeping 
\texttt{ControlResourceSet.controlResourceSetId} unchanged. 
The edited message re-encodes successfully with 
\texttt{pycrate}, remains structurally acceptable, 
and isolates a single semantic violation for attribution.

\subsection{Test Case Execution}

The final phase executes generated test cases on both open-source and commercial 5G UE implementations. 
On open-source UEs, we instrument the program with GDB~\cite{gdb} to capture assertion failures, backtraces, and 
unique crash sites. Each re-encoded test case is delivered via the simulated interface, allowing 
fine-grained triage of violated constraints and their code-level consequences. On commercial smartphones, test cases are delivered over the air through our SDR-based testbed. We monitor the attachment status and system logs via ADB, and 
identify device-level faults such as modem crashes, resets, or persistent misconfigurations. 
This two-path execution strategy ensures that every constraint violation can be observed both 
at the implementation level (open-source UEs) and in real-world devices (OTA testing).
\section{Implementation}

\noindent\textbf{Testbed and Injection Point.}
We deploy a 5G SA testbed where both a simulated UE and commercial smartphones receive mutated downlink RRC messages from the same gNB instance. Concretely, we port the 5Ghoul downlink hook interface from an older OpenAirInterface (OAI) release to the 2025 line of OAI and reuse it to intercept and modify outbound control messages before transmission. The hook is inserted after message construction and before lower–layer scheduling so that registration and connection procedures remain unmodified for both the OAI UE and commercial devices. All seed messages are obtained directly from the unmodified OAI gNB. \looseness=-1

\noindent\textbf{Decoding, Flattening, and Re-encoding.}
All target messages are decoded with pycrate\cite{pycrate} using its RRC ASN.1 library. After decoding, we materialize a flat table of minimally mutable leaf fields and keep the full structural metadata of the original message. Mutations are applied to the flat representation, then the preserved structure is used to reconstruct the hierarchical object and produce a valid Unaligned PER binary. This separation lets us reason at the field level while guaranteeing syntactic compliance on re–encoding.

\noindent\textbf{Schema-driven Probes.}
Value range and presence tests are produced deterministically from TS~38.331. For numeric domains, we instantiate just outside the declared bounds. 
For presence, we remove optional fields using ASN1VE and rebuild the remainder of the message. These cases do not rely on the language model.

\noindent\textbf{LLM-based Constraint Extraction.} We employ GPT-4o~\cite{openai2023gpt4} via the OpenAI API for inferring cross-field dependencies from natural-language specifications. The model is configured with \texttt{temperature=0.0} to ensure stable and reproducible outputs. For LLM-based constraint extraction, the pipeline processed 11{,}005 field pairs in the RRC analysis: 1{,}482 intra-IE pairs and 9{,}523 inter-IE pairs. This yielded 217 candidate DSL rules (66 intra-IE, 151 inter-IE) from the LLM, which were subsequently validated through deterministic domain and satisfiability checks (\S4.2.3). The complete extraction consumed approximately 36.43 million tokens (35.09M input, 1.34M output) over 1.38 hours of processing time.


\section{Evaluation}

\subsection{Experimental Setup}\label{testsetup}

All experiments run on a single workstation with an Intel Core i5-12400 processor (6 cores, 12 threads), 48GB RAM, and Ubuntu 22.04. LLM-based constraint extraction uses the OpenAI API remotely. We illustrate our over-the-air evaluation setup in Appendix Figure~\ref{fig:testbed_setup}. Radio transmission uses a USRP B210 software-defined radio. All experiments are conducted in a laboratory without external cellular coverage to isolate the test network and nearby devices. Commercial phones are configured to PLMN MCC~001/MNC~01 so that they camp on the experiment cell.

\noindent\textbf{Workload and Oracle.}
We evaluated our constraint-guided testing in two settings: commercial smartphones over the air and an open-source UE under simulation.
We focus our experiments on \texttt{RRCSetup}. This message precedes authentication and ciphering; it carries a rich set of configuration information elements that initialize multiple subsystems, and is therefore a natural and high-leverage vehicle to exercise semantic constraints.
A phone run is labeled as a failure if \texttt{adb logcat} shows a modem crash or restart, the data path freezes, or the device cannot attach to or remain connected to the test cell.
An OAI UE run is labeled as a failure if the process crashes or asserts, and we require a reproducible backtrace with a faulting code site.

\subsection{Commercial Smartphone Vulnerability Discovery}\label{sec:phone-vuln}

We test eight commercially available 5G smartphones that cover four chipset families. Device behavior is monitored with Android logcat. We check for modem crash indicators, loss of connectivity to the test cell, and other error signatures. Detailed information on the tested devices and their baseband versions is summarized in Appendix Table~\ref{tbl:device_info}.

Table~\ref{tab:smartphone-vuln} summarizes each finding by chipset vendor, the violated constraint class, the modified Information Element, a short log excerpt indicating the failure mode, and the disclosure status. We highlight three representative observations from our sample rather than drawing vendor-wide conclusions. On MediaTek devices, value range violations yielded multiple distinct crash points across different IEs. Following the mutation procedure in §4.3, we modified individual fields to out-of-range values and re-encoded the complete message. While these messages remained syntactically valid—enabling Wireshark and ASN1VE to parse them with only semantic warnings at the modified fields—the baseband implementations failed to handle the constraint violations, leading \texttt{adb logcat} to report fatal error codes. The affected devices could not reattach until a reboot, constituting a denial of service. 

\begin{figure}[t]
    \centering    
    \includegraphics[width=0.95\columnwidth]{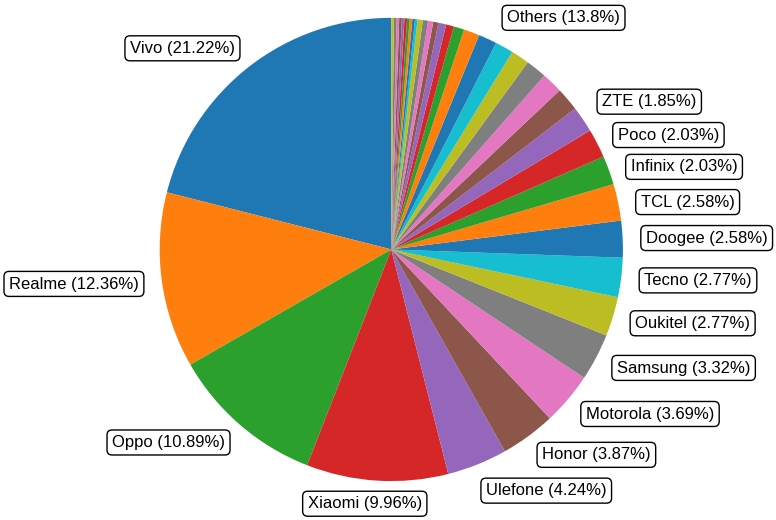}
    \caption{Smartphone Brands Affected by Confirmed Flaws}
    \label{fig:affected_smartphones}
\end{figure}

As of submission, MediaTek has assigned three high-severity CVE IDs. Qualcomm has confirmed several issues and granted a reward for one issue. Other entries remain under review as indicated in the disclosure column. We followed responsible disclosure and limited our claims to denial-of-service, based on logcat evidence and loss of connectivity. These confirmed flaws impact \textbf{64} chipset models and over \textbf{542} commercially available smartphone models~\cite{garbelini2023_5ghoul,kimovil2025,mediatek2025,qualcomm2025}, with additional details provided in  Figure~\ref{fig:affected_smartphones}.

\begin{table*}[t]
\centering
\setlength{\tabcolsep}{5.5pt}
\renewcommand{\arraystretch}{1.2}
\caption{Commercial Smartphone Vulnerability Discovery}
\label{tab:smartphone-vuln}
\footnotesize
\begin{tabular}{
    l
    l
    >{\raggedright\arraybackslash}p{2.6cm}
    >{\raggedright\arraybackslash}p{7.6cm}
    l
}
\toprule
\textbf{Vendor} & \textbf{Constraints } & \textbf{Affected IE}                                                             & \multicolumn{1}{c}{\textbf{Crash Log Snippet}}                                                                                                         & \multicolumn{1}{c}{\textbf{Disclosure}}                          \\ 
\midrule
\textbf{MediaTek}       & Value Range                    & multiple IEs                                                                        & Fatal Error  \& Fatal error err\_code1:0x0000001D                                                                                                      & \begin{tabular}[c]{@{}l@{}}\textbf{CVE-2025-20644}\\ Patched\end{tabular} \\ \hline
\textbf{MediaTek}       & Intra-IE                       & csi-MeasConfig                                                                            & cause:(VSONIC0) {[}Fatal error{]} err\_code1:0x00000020                                                                                                & \begin{tabular}[c]{@{}l@{}}\textbf{CVE-2025-20703}\\ Patched\end{tabular} \\ \hline
\textbf{MediaTek}       & Intra-IE                       & ServingCellConfig                                                                              & \begin{tabular}[c]{@{}l@{}}{[}ASSERT{]} file:mcu/l1/mml1/mml1\_rf/src/mmrf/gen97/\\ mml1\_rf\_ucnt.c line:798\end{tabular}                             & \begin{tabular}[c]{@{}l@{}}\textbf{CVE-2025-20666}\\ Patched\end{tabular} \\ \hline
\textbf{Qualcomm}       & Inter-IE                       & tag-Config                                                                         & modem nrul\_gm\_ulm.c:2090:NRUL\_IUSS: Assertion                                                                                                       & \textbf{Confirmed}                                                        \\ \hline
\textbf{Qualcomm}       & Intra-IE                       & srs-Config                                                                         & modem nrfw\_sys\_ul\_ch.c:1860:Assertion                                                                                                               & \begin{tabular}[c]{@{}l@{}}\textbf{Rewarded}\\ \end{tabular}   \\ \hline
\textbf{Qualcomm}       & Inter-IE                       & \begin{tabular}[c]{@{}l@{}}ControlResourceSet\\SearchSpace\end{tabular} & \begin{tabular}[c]{@{}l@{}}subsys\_err\_fatal\_intr\_handler(): Fatal error on modem!\\ modem nr\_rx\_ctl\_cch.c:1116:NR\_RX\_IUSS: ERROR\end{tabular} & Under Review                                                     \\ \hline
\textbf{Qualcomm}       & Intra-IE                       & srs-Config                                                                         & \begin{tabular}[c]{@{}l@{}}subsys\_err\_fatal\_intr\_handler(): Fatal error on modem! \\ modem nr5g\_ml1\_ulrm.c:1565:Assert 0 failed\end{tabular}     & Under Review                                                     \\ \hline
\bottomrule
\end{tabular}
\end{table*}

For completeness, we note that several additional findings were also acknowledged by vendors but classified as duplicates of prior reports; these duplicates are excluded from Table~\ref{tab:smartphone-vuln} and from all counts.

\noindent\textbf{Security Implications.}
Because all affected vulnerabilities manifest during the pre-authentication stage, an adversary can exploit them without any valid credentials or cryptographic material. A rogue gNB, implemented with commodity SDR hardware and open-source stacks, can deterministically deliver crafted yet syntactically valid RRC messages that trigger baseband crashes or attach failures. This capability enables a practical adversary to achieve persistent denial-of-service conditions against nearby devices. In many cases, the affected smartphones remain unable to reconnect until a manual reboot, effectively rendering the user unreachable. From the user’s perspective, this translates into a loss of service availability—calls, data sessions, and emergency communications are disrupted, and the device must be restarted to recover.

\subsection{OAI UE Vulnerability Discovery}\label{sec:oai-vuln}

\begin{table*}[t]
\centering
\setlength{\tabcolsep}{5.5pt}
\renewcommand{\arraystretch}{1.2}
\caption{OAI UE Vulnerability Discovery}
\label{tab:OAIUE-vuln}
\footnotesize
\begin{tabular}{
    >{\raggedright\arraybackslash}p{2.5cm}  
    l  
    >{\raggedright\arraybackslash}p{3.0cm}  
    l  
}
\toprule
\textbf{Constraints} & \multicolumn{1}{c}{\textbf{Affected IE}} & \multicolumn{1}{c}{\textbf{UE Crash Reason}} & \multicolumn{1}{c}{\textbf{Crash Line of Code}}             \\ \midrule
Value Range          & csi-MeasConfig                             & Segmentation Fault                                & openair2/LAYER2/NR\_MAC\_COMMON/nr\_mac\_common.c:5122      \\ \hline
Value Range          & pucch-Config                               & Assertion                                    & openair2/LAYER2/NR\_MAC\_UE/nr\_ue\_procedures.c:2223       \\ \hline
Presence             & srs-Config                                 & Segmentation Fault                                & openair2/LAYER2/NR\_MAC\_UE/nr\_ue\_power\_procedures.c:498 \\ \hline
Presence             & pdsch-servingcellconfig                    & Assertion                                    & openair2/LAYER2/NR\_MAC\_UE/nr\_ue\_procedures.c:1344       \\ \hline
Intra-IE             & srs-Config                                 & Segmentation Fault                                & openair1/PHY/NR\_UE\_TRANSPORT/srs\_modulation\_nr.c:383    \\ \hline
Intra-IE             & csi-MeasConfig                             & Assertion                                    & openair2/LAYER2/NR\_MAC\_COMMON/nr\_mac\_common.c:5109      \\ \hline
Inter-IE             & csi-MeasConfig                               & Assertion                                & openair2/LAYER2/NR\_MAC\_COMMON/nr\_mac\_common.c:5139                          \\ \hline
Inter-IE             & pdcch-Config                               & Assertion                                    & openair2/LAYER2/NR\_MAC\_UE/nr\_ra\_procedures.c:79         \\ \bottomrule
\end{tabular}
\vspace{0.3em}

{\footnotesize\textit{* This table lists 8 representative cases; the complete set of 29 crash sites is available in the public repository.}}
\end{table*}

We exercised all four constraint classes and delivered the mutated messages to the OAI UE, and used automated GDB backtraces to localize failures. In total, we identified 29 unique crash sites triggered when the UE receives our test messages. Table~\ref{tab:OAIUE-vuln} lists the cases by constraint class, vulnerable IE, failure type, and first faulting line from GDB.

A value range violation in \texttt{csi-MeasConfig} produced a segmentation fault in the MAC common layer, 
while a range tweak in \texttt{pucch-Config} triggered an assertion in the UE MAC procedures. 
Removing an optional field in \texttt{srs-Config} led to a null pointer dereference in UE power control, 
and an intra-IE inconsistency in \texttt{srs-Config} caused a crash in the PHY SRS path. 
For inter-IE dependencies, a mismatch in PDCCH configuration produced an assertion in the random access procedures, 
and another input induced a segmentation fault along the RF simulator path. 
These examples show that constraint violations map to concrete and repeatable failure sites across MAC, PHY, and simulation components.

Beyond the representative examples, we further summarize how distinct crash sites are distributed across constraint classes.
Table~\ref{tab:crash_sites_by_constraint} reports the number of unique crash sites triggered by each constraint class.

\begin{table}[t]
\centering
\caption{OAI UE Crash Sites Triggered by Different Constraint Classes (counts may overlap)}
\label{tab:crash_sites_by_constraint}
\vspace{2ex} 
\begin{tabular}{l@{\hspace{0.3em}}c@{\hspace{0.3em}}c@{\hspace{0.3em}}c@{\hspace{0.3em}}c}
\toprule
\textbf{Constraint Class} & \textbf{Value} & \textbf{Presence} & \textbf{Intra-IE} & \textbf{Inter-IE} \\
\midrule
\textbf{Distinct Crash Sites} & 16 & 5 & 8 & 3 \\
\bottomrule
\end{tabular}
\end{table}

\noindent\textbf{Responsible Coordination.}
OAI is an actively maintained open-source project with frequent updates. 
Accordingly, we prioritized responsible coordination with the OAI developers rather than pursuing CVE assignment and shared relevant information to support issue resolution. 
As of submission, \textbf{four} fixes have been merged in response to our reports, and one issue was assigned a CVE (\textbf{CVE-2025-63356}).\looseness=-1

\subsection{Constraint-guided Testing vs Enumeration Baselines}
We compare our constraint-guided testing with an exhaustive pairwise enumeration baseline towards OAI UE in the simulation environment. Both methods operate on \texttt{RRCSetup} and are evaluated on the OAI user equipment under simulation. The baseline exhaustively iterates over the ASN.1 domains of the two target fields while keeping all other fields unchanged. Our method first extracts field-level constraints from the specifications and then generates a small set of targeted edits that exercise those constraints.

\noindent\textbf{Metrics.}
Both methods reencode the modified messages with \texttt{pycrate}, follow the same delivery path, and use the same oracle. A run is labeled as a failure only when the OAI UE process crashes or asserts with a reproducible backtrace. Following \textsc{5Ghoul}, each RRC connection test is monitored for 45 seconds~\cite{garbelini2023_5ghoul}. To keep the experiment manageable, we uniformly sample 6{,}000 field pairs from all \(\binom{939}{2}=440{,}391\) possible pairs in a seed \texttt{RRCSetup} message generated by the OAI gNB.

\noindent\textbf{Results.}
Under this setup, constraint-guided testing generates substantially fewer inputs than enumeration while still exposing meaningful failures. Specifically, constraint-guided testing generates 1{,}458 inputs across the sampled pairs, whereas enumeration produces 30{,}600 inputs over the sampled field pairs. The total elapsed time to complete each batch is 0.76 days and 15.94 days, respectively. Under the same per-pair budget, fully enumerating all \(\binom{939}{2}\) pairs would require approximately \(2.25\times10^{6}\) runs and about \(1.17\times10^{3}\) days.

The failures uncovered by constraint-guided testing are easy to attribute to explicit specification constraints. In contrast, enumeration yields many more inputs, and three of its failures arise from not implemented or not supported assertions that act as development kill switches rather than security-relevant issues. Presence constraints are a particular strength of our method, as they rely on need code semantics that are not visible to pure enumeration.


\noindent\textbf{Takeaway.}
While exhaustive enumeration can surface a broad range of behaviors under simulation, its input volume and execution cost make it impractical to scale. In contrast, constraint-guided testing produces smaller, more focused input sets that are better aligned with exercising specification-relevant failure paths. These properties make constraint-guided testing more suitable for realistic OTA campaigns. To further contextualize this setting, we additionally evaluate a grammar-aware but semantics-agnostic mutation baseline under a fixed OTA execution budget; details are provided in Appendix~\ref{sec:baseline}.


\subsection{Sanity check: Evidence vs Schema in Rule Extraction}
We study how the available evidence affects LLM extraction of \emph{intra-IE} field-dependency rules. To isolate the effect of evidence, the model, prompt, temperature, and post-processing are fixed; only the provided evidence varies. We restrict the evaluation to pairs of fields within the same IE to avoid cross-IE aliasing and identifier confounds. We compare three settings: \emph{Full Evidence} (TS~38.331 ASN.1, field descriptions, and cross-document snippets from TS~38.213/38.214/38.321/38.322), \emph{No CrossDocs} (TS~38.331 ASN.1 and field descriptions only), and \emph{ASN.1-only} (the ASN.1 block of the target IE only). We report the \emph{rule output rate} as defined below.

\noindent\textbf{Metric.}
We evaluate on a fixed set of 1,482 \emph{intra-IE} field pairs~(two fields within the same IE) preselected from specification sentences that co-mention the targets and contain deontic cues~(e.g., ``shall'', ``only if''). Our metric is the \emph{rule output rate}: the number of pairs for which the model returns a DSL rule~(rather than \texttt{NO\_RULE}). A candidate rule is counted only if it (i) includes a verbatim quote from the provided snippets, and (ii) uses values or ranges that appear in the given ASN.1. Pairs that fail either gate are mapped to \texttt{NO\_RULE}. We do not evaluate crash triggering here; the goal is to isolate how evidence affects the ability to derive an executable, source-backed rule. This intra-IE focus avoids confounding factors from cross-IE aliasing and identifiers.

\noindent\textbf{Results.}
On the same 1,482 intra-IE field pairs extracted by our pipeline, we evaluate the effect of different evidence configurations.
\emph{Full Evidence} (TS~38.331 ASN.1 and field descriptions combined with cross-document prose from the TS~38.2xx family) yields 57 valid DSL rules. 
\emph{No CrossDocs}~(TS~38.331 only) yields 9 rules, while \emph{ASN.1-only} yields no rules, as no normative sentence co-mentions both fields and the citation gate fails. 
Overall, 86.4\% (57/66) of the valid intra-IE DSL rules depend on cross-document prose, indicating that behavioral specifications outside TS~38.331 are essential for extracting executable intra-IE semantic constraints.
This ablation focuses on intra-IE pairs; inter-IE dependencies are primarily specified in TS~38.213/38.214 prose, and their necessity is illustrated by the inter-IE case studies in \S\ref{sec:case-studies}.

\noindent\textbf{Takeaway.}
Combining schema and prose is essential. TS~38.331 alone gives precise domains but misses many field–to–field relations; cross–document prose supplies those relations but needs the schema to anchor values. Using both yields the highest rule output rate and produces executable rules that cleanly drive message edits.

\section{Case Studies}
\label{sec:case-studies}

Our constraint violation case studies span open-source UE and commercial 5G smartphones. On commercial devices, the baseband is a black box, so we rely on Android \texttt{logcat} and connectivity symptoms rather than source traces. For the open source OAI UE, when a crash occurs, we attach GDB to collect a backtrace and identify the faulting site.

\noindent\textbf{Field Value Range Constraints Violation. } The target field \texttt{startPosition} has an ASN.1 domain of \textbf{\texttt{0..5}} in the 3GPP TS 38.331 Page 827. We construct test cases that change only this field while leaving the message structure and all other values intact. Two out of range assignments are used, \textbf{-1} and \textbf{6}. \looseness=-1



\begin{figure}[h]
    \centering    
    \includegraphics[width=1.0\columnwidth]{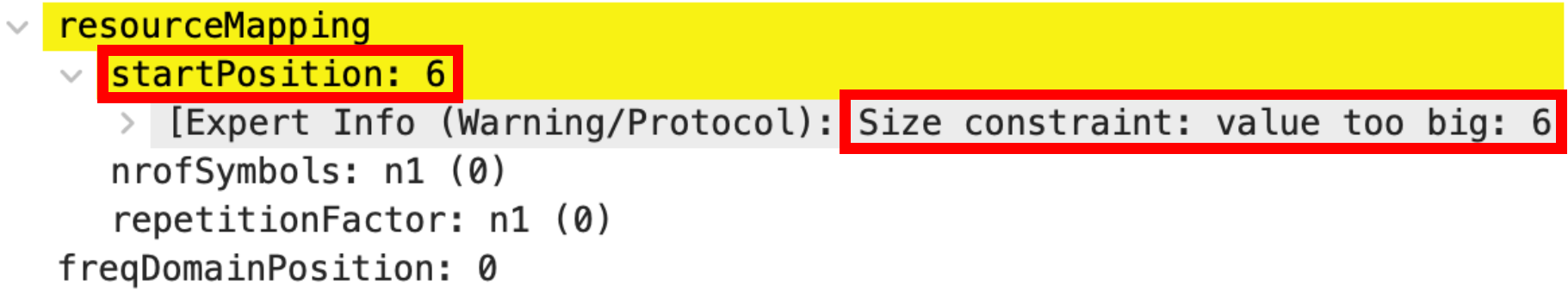}
    \caption{Value Range Constraint Violation}
    \label{fig:value_violation}
\end{figure}

\noindent\textbf{Outcome and Root Cause.}
Both out-of-range values were accepted by the pycrate ASN.1 encoder despite safety flags being enabled \texttt{\_SAFE\_VAL}, \texttt{\_SAFE\_BND}, \texttt{\_SAFE\_BNDTAB} = \texttt{True}. The effective guard for this field was inactive because the constraint object was unset \_const\_val = \texttt{None}, which bypassed bound checking and produced a syntactically well-formed message. The OAI UE decoded the message but performed no post-decode range validation for \texttt{startPosition}, allowing the invalid value to propagate into later computations and causing a crash. Proper hardening would include validating numeric fields against their declared domains after decoding, applying defensive bounds on derived indices and lengths, and handling errors without terminating the process. The security impact is a practical denial of service condition that an adversary can trigger by injecting a control plane message whose encoding is accepted while its numeric content violates the specification.

\noindent\textbf{Commercial Smartphone Evidence.}
We applied the value range violation to commercial smartphones. For a chosen IE, we set one field to an out-of-range value, reencoded only that IE, and spliced the bytes back into an otherwise unchanged downlink message. Wireshark and ASN1VE could still parse the packet, but flagged a warning at the modified IE. On two devices with MediaTek chipsets, this input exposed multiple distinct crash points across different IEs. Android \texttt{logcat} recorded modem exception traces and abort messages consistent with assertion failures and memory corruption. After receiving the input, the smartphone failed to attach to the test gNB and did not reconnect until a manual reboot, which constitutes a denial of service. We disclosed the issue to the MediaTek Security Team; it was triaged as high severity and assigned \textbf{CVE-2025-20644}. This subcase shows that field-level violations of value range constraints can cause critical failures in production stacks.

\noindent\textbf{Field Presence Constraints Violation.} We treat presence semantics as a deterministic class of constraints derived from the ``need code’’ annotations in TS~38.331. As a focused experiment, we selected one IE optional field \texttt{spatialRelationInfo}, removed only this IE field from an otherwise valid downlink RRC message, and sent it to the OAI UE. On the OAI UE, receiving a message with this single optional field removed consistently caused an immediate crash with one hundred percent reproduction, even though the remainder of the message remained syntactically valid.

\begin{figure}[h]
    \centering    
    \includegraphics[width=0.70\columnwidth]{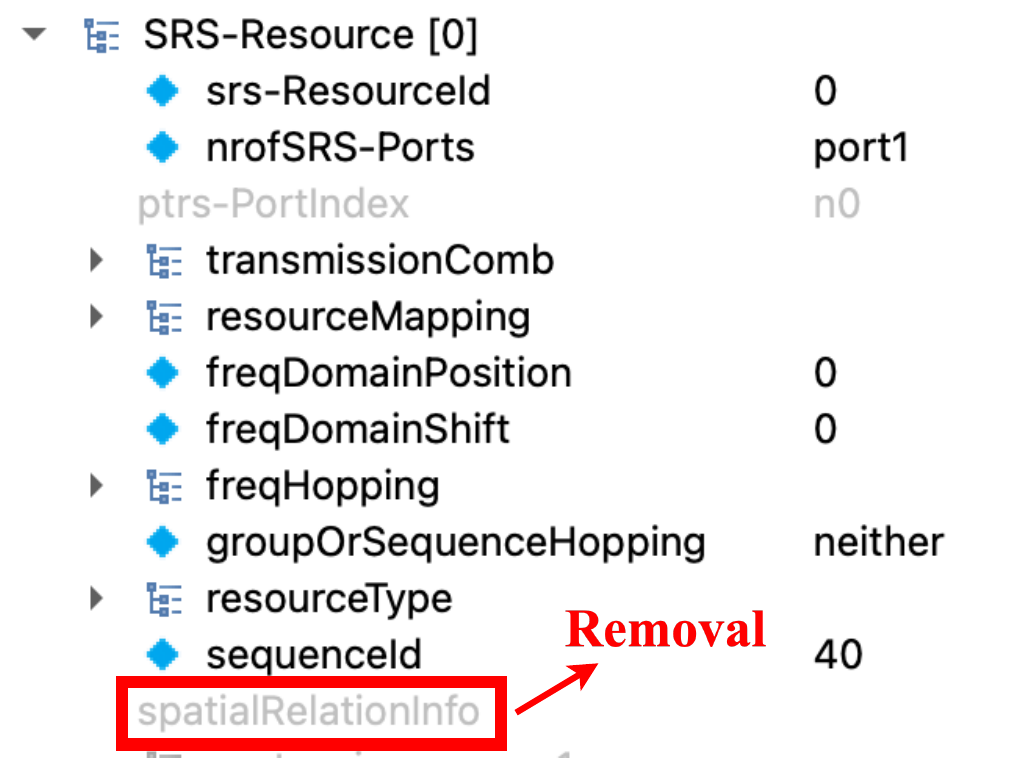}
    \caption{Presence Constraint Violation Example}
    \label{fig:presence_case}
\end{figure}

Further inspection indicates that the absence of \texttt{spatialRelationInfo} leads to an inconsistent internal configuration state during subsequent processing, ultimately causing the UE to fail.
This behavior suggests that the implementation implicitly relies on the presence semantics specified in the standard, but does not defensively handle violations after decoding.
The result is a low-cost denial of service triggered by removing a single optional field, highlighting the security impact of presence constraints.

\begin{figure}[h]
    \centering    
    \includegraphics[width=1.0\columnwidth]{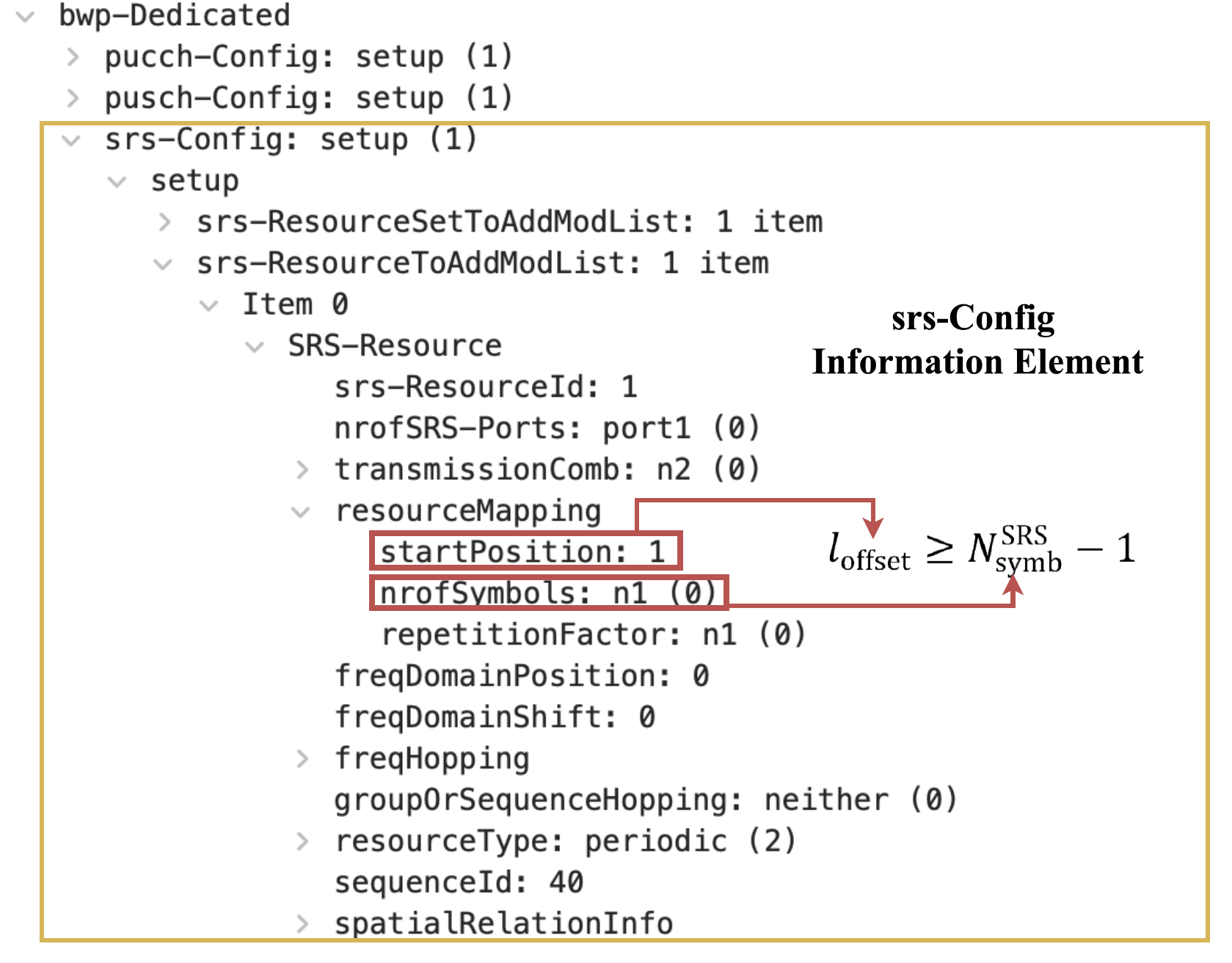}
    \caption{Intra-IE Field Constraint Example}
    \label{fig:Intra_IE_example}
\end{figure}

\noindent\textbf{Intra-IE Field Dependency Constraint Violation.}
Within the \texttt{srs-Config} IE, \texttt{nrofSymbols} (number of consecutive symbols) and \texttt{startPosition} (the starting symbol counted backwards from the end of the slot) are jointly constrained by TS~38.211~\S6.4.1.4.1.
Letting $N^{\text{SRS}}_{\text{symb}}$ denote \texttt{nrofSymbols} and $l_{\text{offset}}$ denote \texttt{startPosition}, the specification states that
$l_{\text{offset}} \geq N^{\text{SRS}}_{\text{symb}} - 1$,
which ensures that the derived time-domain mapping remains consistent with the slot structure (see also the TS~38.331 field description of \texttt{resourceMapping}).
Violating this constraint can make the derived time-domain mapping inconsistent with the slot structure, which in turn may exercise unsafe receiver behaviors.
\ProjectName{} normalizes this requirement into an executable DSL constraint that captures the semantic dependency between the two fields.
Figure~\ref{fig:Intra_IE_example} illustrates the relevant fields and their relationship within the \texttt{srs-Config} IE.

To construct test cases, \ProjectName{} deliberately violates this extracted DSL constraint while preserving overall message well-formedness. In our testbed, we increased \texttt{nrofSymbols} from \texttt{n1} to \texttt{n4} while leaving the remaining \texttt{SRS-Resource} configuration unchanged, thereby breaking the semantic relationship between \texttt{nrofSymbols} and \texttt{startPosition} and resulting in an invalid SRS configuration. The resulting RRC message consistently triggered a UE crash upon reception, demonstrating that targeted violations of intra-IE semantic constraints can exercise meaningful failure paths and lead to denial of service.


\noindent\textbf{Commercial Smartphone Evidence.}
On a commercial 5G smartphone, we targeted another intra-IE dependency inside \texttt{srs-Config} and broke it with a minimal edit while keeping the message otherwise valid. When the smartphone received this falsified RRC message, the device reproducibly entered a modem crash state, with \texttt{logcat} showing modem exception traces, and it could not reattach until a reboot. Representative cases are summarized in Table~\ref{tab:smartphone-vuln}.

\noindent\textbf{Inter-IE Field Dependency Constraints Violation.} An illustrative Inter-IE constraint arises in PDCCH configuration: 3GPP TS38.213 \S10.1 requires that each \texttt{SearchSpace} be associated with a \texttt{ControlResourceSet} via \texttt{controlResourceSetId} \footnote{CORESET is the common abbreviation for \emph{ControlResourceSet} as used in 3GPP specifications.}.

\begin{tcolorbox}[breakable]

\textbf{\textcolor{blue!50!black}{[3GPP TS 38.213 Page 144]}}:

...an association between the \textbf{search space} set and a \textbf{CORESET} by \textbf{controlResourceSetId} or by controlResourceSetId-v1610...

\end{tcolorbox}

\begin{figure}[h]
    \centering    
    \includegraphics[width=1.0\columnwidth]{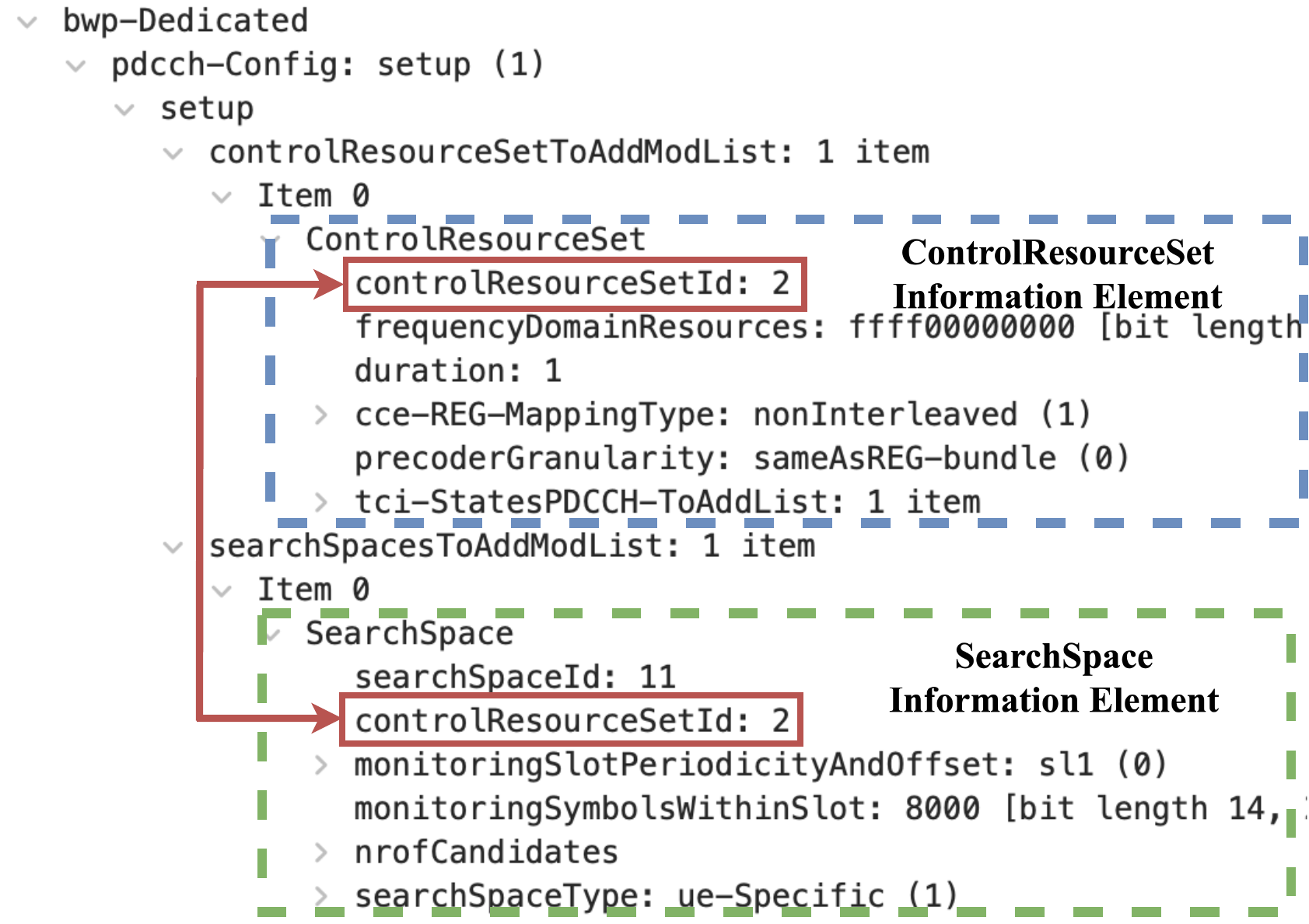}
    \caption{Inter-IE Field Constraint Example.}
    \label{fig:Inter_IE_example}
\end{figure}

In our experiment towards testing OAI UE, the original RRC message ensured consistency by setting both the \texttt{controlResourceSetId} from \texttt{ControlResourceSet} IE and the corresponding \texttt{controlResourceSetId} from the \texttt{SearchSpace} IE to the same value (e.g.,~2). We intentionally broke this Inter-IE linkage by changing the \texttt{controlResourceSetId} from \texttt{SearchSpace} IE from 2 to 11 while leaving the \texttt{controlResourceSetId} from \texttt{ControlResourceSet} IE at 2. This mismatch violates the above normative association and reproducibly caused the OAI UE to crash when receiving this falsified RRC message.

\noindent\textbf{Commercial Smartphone Evidence.}
On a commercial 5G smartphone, configuring \texttt{SearchSpace} and \texttt{ControlResourceSet} with inconsistent
\texttt{controlResourceSetId} values consistently triggered scheduler receive path assertions in \texttt{logcat}. 
This behavior indicates a violation of internal linkage assumptions between information elements, causing the device to fail in handling otherwise valid-looking RRC messages. 
The observed failure is therefore consistent with an Inter-IE dependency bug, as summarized in Table~\ref{tab:smartphone-vuln}.

\section{Discussion and Limitations}

\subsection{Constraint Modeling Scope and Completeness}
\noindent\textbf{Constraint Taxonomy and Coverage.} 
The space of semantic constraints in 5G messages is large and often ambiguously specified. 
Our four-class taxonomy—field value ranges, field presence conditions, intra-IE field dependencies, and inter-IE field consistencies—distills recurring patterns in 3GPP specifications and aligns with baseband implementation logic. 
Empirically, it enables targeted test generation and crash discovery on both open-source and commercial UEs, indicating that these constraints capture critical aspects of real implementations. 
While additional types, such as cross-protocol or temporal dependencies, may exist, the framework is modular and can be extended to incorporate them.

In particular, some specification constraints may be expressed implicitly at the procedural or behavioral level, rather than as explicit field-to-field relations. Such dependencies are difficult to capture with our current evidence-bound extraction strategy, which focuses on constraints that can be localized to individual messages or IEs. As a result, a small number of constraints governed by a broader procedural context may not be identified.

\noindent\textbf{DSL Expressiveness.}
The current DSL captures field-level constraints within individual messages but does not express \emph{capability-based rules} (e.g., constraints conditioned on prior capability negotiation), \emph{state-based rules} (e.g., field requirements dependent on protocol state~\cite{ranjbar2025stateful}), or \emph{temporal constraints} (e.g., inter-message timing and ordering requirements). 
As a result, \ProjectName{} does not explicitly represent or test 5G behaviors whose validity depends on cross-message context, such as capability-negotiated feature activation, protocol-state–dependent procedures, or timing- and ordering-sensitive requirements governed by protocol timers.

In addition, some constraints are specified through complex tables or mathematical expressions in 3GPP documents, which are not always reliably parsed. While we use OCR-based techniques to improve extraction coverage, fully reliable extraction is not guaranteed due to formatting variability, and a small number of constraints may therefore be missed.

\subsection{Applicability to Other Protocols}

While our end-to-end evaluation focuses on RRC, we conduct a preliminary study to demonstrate that \ProjectName{}'s constraint extraction generalizes to other ASN.1-based 5G protocols. We applied it to NGAP (TS 38.413) and F1AP~(TS 38.473), which govern core-to-RAN and intra-gNB communication, respectively. Table~\ref{tab:protocol_generality} summarizes the extracted constraints. 

\begin{table}[t]
\centering
\caption{Constraint Extraction Results for NGAP and F1AP}
\label{tab:protocol_generality}
\vspace{2ex} 
\begin{tabular}{l@{\hspace{1em}}c@{\hspace{1em}}c@{\hspace{1em}}c@{\hspace{1em}}c}
\toprule
\textbf{Protocol} & \textbf{Value} & \textbf{Presence} & \textbf{Intra-IE} & \textbf{Inter-IE} \\
\midrule
NGAP  & 92  & 174 & 55 & 18 \\
F1AP  & 277 & 211 & 45 & 11 \\
\bottomrule
\end{tabular}
\end{table}

These results demonstrate that \ProjectName{} successfully extracts semantic constraints from diverse 5G control-plane protocols. However, comprehensive evaluation of these protocols remains future work. NGAP testing would require a full 5G core deployment with N2 interface instrumentation, while F1AP testing demands multi-node gNB configurations with CU-DU split architectures. Given appropriate testbeds and seed messages, \ProjectName{}'s test case generation and mutation engines (§4.3) could be applied to these protocols following the same methodology as RRC, as the extraction and mutation logic remain protocol-agnostic. 

Compared to RRC, other protocols exhibit distinct characteristics that may introduce additional challenges when applying \ProjectName{}. For example, NGAP and F1AP involve richer procedural context and more complex multi-entity interactions than RRC. More generally, \ProjectName{} is applicable to any protocol that defines message structures in ASN.1 and specifies semantic constraints through normative natural-language text. For such protocols, adapting \ProjectName{} primarily requires binding the extracted constraints to the appropriate message scope and execution context, while the evidence extraction, DSL representation, and mutation logic remain largely reusable.

\subsection{LLM Hallucination Mitigation}

We restrict LLM use to intra- and inter-IE field dependencies; numeric ranges and presence rules are deterministically derived from TS~38.331. For each target pair, we assemble an evidence slice with the ASN.1 block, field names, and normative sentences containing deontic cues (e.g., ``shall'' or ``must''). The model is instructed to reason only over this slice and return \texttt{NO\_RULE} if support is lacking. Accepted rules must cite the source text and are normalized into DSL predicates. Each candidate rule then passes deterministic gates to check scope, citations, and domain validity, with conflicts or release-specific clauses conservatively filtered. Our ablation results show that cross-document prose is often necessary for extracting executable rules: removing it sharply reduces extracted rules, while ASN.1 alone yields none. This indicates that our gates suppress unsupported inferences and that many useful relations exist only in natural-language standards. Although some constraints could be hand-scripted, cross-document dependencies are too numerous to enumerate manually, and the LLM recovers substantially more rules than deterministic parsing alone.

Residual risks remain, such as version drift or ambiguous naming, which we mitigate through version pinning, alias normalization, and strict admission criteria. Resulting tests apply localized edits so that each observed effect can be attributed to one violated rule.
\section{Related Work}
\noindent\textbf{Specification-Guided Analysis}. Prior efforts detect contradictions or ambiguities in specifications using natural language and machine learning~\cite{rahman2024cellularlint, chen2022seeing}, or interpret deviant behaviors post hoc through models derived from informal descriptions or conformance documents~\cite{hussain2021noncompliance, karim2021prochecker, tu2024logic}. Bookworm Game~\cite{chen2021bookworm} applies natural language processing to extract test objectives from LTE specifications, identifying procedural constraints and state-machine requirements from normative text for testing. Recent systems leverage specification-derived structures for test synthesis and semantic reasoning~\cite{chen2023sherlock, karim2023spec5g}, yet rarely capture field-level semantic dependencies. In contrast, our work systematically mines and formalizes cross-field constraints from specifications to guide baseband vulnerability discovery.\looseness=-1

\noindent\textbf{Uncovering 5G Implementation Flaws}.
Reverse engineering exposes memory and logic flaws but is limited by firmware access and manual effort~\cite{golde2016breaking,grassi2018exploitation}; emulation-based testing identifies inconsistencies via conformance checks, state models, or differential analysis without requiring physical devices~\cite{hussain2021noncompliance,tu2024logic}; and OTA fuzzing exploits unauthenticated RRC phases with crafted messages to uncover parsing and assertion bugs~\cite{park2022doltest,hoang2025llfuzz}. 
Recent work has begun exploring LLM-guided fuzzing. For example, ChatAFL~\cite{meng2024large} utilizes large language models to generate protocol seeds and guide mutation strategies, thereby improving code coverage in general network protocols. 
While these methods reveal syntactic issues such as missing fields or corrupted ASN.1 structures, they rarely capture semantic inconsistencies arising from inter-field dependencies.\looseness=-1

\noindent\textbf{Testing Other 5G Network Components}.
Beyond UE testing, recent work has explored the security of 5G core networks. RANsacked~\cite{bennett2024ransacked} proposes ASNFuzzGen, a structure-aware framework for reconstructing ASN.1 formats from protocol specifications, enabling domain-informed fuzzing of LTE and 5G RAN-core interfaces with multi-threaded coverage collection. The framework focuses on improving input validity and code coverage through grammar-guided mutations and instrumentation-based feedback. CORECRISIS~\cite{dongcorecrisis} applies FSM learning to drive NAS-layer fuzzing. These approaches add structure and state awareness but still rely on large-scale fuzzing, which is impractical in black-box OTA settings due to reboots and SDR delays. Our work instead extracts semantic constraints from specifications to generate targeted tests, improving efficiency and scalability for both UE and core protocols.
\section{Conclusion}


This work shows that 3GPP specification text can be converted into executable rules and targeted tests to uncover logic-level semantic inconsistencies in 5G RRC implementations. \ProjectName{} combines schema-defined constraints from TS~38.331 with cross-field dependencies mined from normative text, expressed in a compact rule language. It uncovered 7 flaws on commercial smartphones with 3 assigned CVEs, and identified 29 crash sites in the open-source OAI UE. These results highlight logic-level semantic inconsistencies as a distinct class of flaws overlooked by existing OTA fuzzers and position \ProjectName{} as a systematic approach to specification-guided testing in cellular systems.

\section{Acknowledgements} 


We thank the anonymous reviewers for their valuable comments.
This material is based upon work supported in part by the National Science Foundation (NSF) under Grant No.~2228617, 2120369, 2129164, 2148374, and 2226339.
Any opinions, findings, and conclusions or recommendations expressed in this material are those of the authors and do not necessarily reflect the views of the NSF.




\section*{Ethical Considerations}
This work studies robustness issues in 5G UE implementations by injecting semantically inconsistent RRC messages during the pre-authentication phase. The experiments directly interact with real protocol stacks and commercial devices, which requires careful control of scope and impact.

\textbf{Experimental Environment.}
All over-the-air experiments were conducted in a controlled laboratory setting using an isolated 5G test network with no connectivity to commercial cellular infrastructure. Only devices owned and operated by the research team were used as test subjects. No third-party devices were affected, and no user traffic or personal data was collected or observed. We did not test live networks or devices outside our control.

\textbf{Impact on Devices and Services.}
Our test cases can trigger denial-of-service conditions, such as modem crashes or attach failures, on affected devices. To limit harm, experiments were performed on dedicated test devices, and failures were recovered through reboot or reflashing. We did not attempt to sustain or amplify outages beyond what was necessary to confirm reproducibility, and we did not evaluate persistence, large-scale deployment, or real-world attack scenarios.

\textbf{Vulnerability Disclosure.}
For vulnerabilities affecting commercial devices, we followed responsible disclosure practices. We reported confirmed issues privately to the corresponding vendors, provided reproducing inputs and diagnostic evidence, and allowed time for investigation and patching prior to publication. Public descriptions in this paper are intentionally high-level and do not include exploit-ready payloads that would lower the barrier for misuse. For issues identified in the OpenAirInterface UE, we reported reproducible crashes and diagnostic traces directly to the maintainers, and several fixes have been merged upstream.

\textbf{Dual-Use Considerations.}
The techniques in this paper demonstrate that semantically inconsistent but standard-compliant RRC messages can cause deterministic failures in UE implementations. While such techniques could be misused to induce denial-of-service, the underlying weaknesses already exist in deployed systems and are reachable through previously studied rogue base station models. We chose to publish our findings to enable vendors and researchers to systematically identify and mitigate these classes of implementation flaws rather than leaving them undocumented.

\textbf{Use of Language Models.}
Large Language Models are used only to assist in extracting semantic constraints from natural-language specifications, not to generate exploits, payloads, or attack strategies. All extracted constraints are validated against specification evidence and ASN.1 domains before being used for test generation. We design our experiments to minimize unintended impact while enabling reproducible evaluation of implementation robustness. We believe that documenting these flaws and the methodology used to uncover them provides practical value for improving the security and reliability of 5G UE implementations.




\section*{Open Science}
We release three artifacts to enable end-to-end reproduction: (A1) a simulation testbed with an OAI gNB where 5Ghoul-style downlink interception hooks are updated to 2025.w5 branch as 5G SA setup; (A2) a constraint-driven toolchain comprising document-slicing and field-pairing scripts, strong-cue filters, one prompt template with batch drivers for evidence-to-DSL synthesis and pycrate-based ASN.1 UPER decode/encode utilities; and (A3) test case and corresponding exploit code that via the OAI gNB interception path and proof-of-concept exploits in the simulation environment. All provided exploit code only works in simulation and is not usable over-the-air. OTA-specific exploits are deliberately withheld to prevent misuse.


\bibliographystyle{plain}
\bibliography{references}
\appendix

\section{Testbed and Tested Devices} 

\begin{figure}[h]
    \centering    
    \includegraphics[width=1.0\columnwidth]{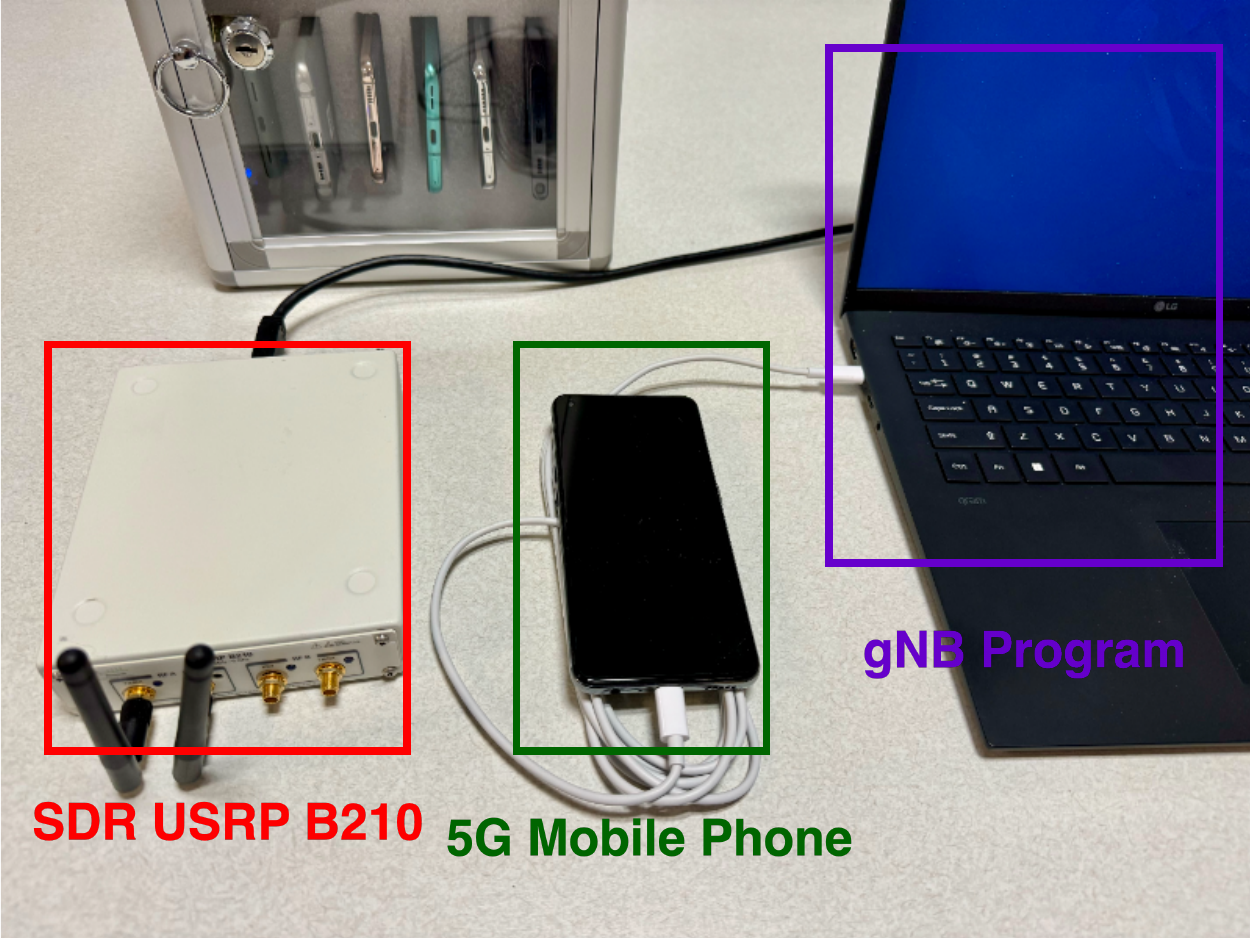}
    \caption{Testbed Setup for Commercial Smartphones.}
    \label{fig:testbed_setup}
\end{figure}

As illustrated in Figure~\ref{fig:testbed_setup}, we employ a USRP B210 as the radio front-end and run the OAI gNB on a dedicated Ubuntu 22.04 laptop. All experiments are carried out in an isolated laboratory environment. Each tested smartphone is equipped with a customized SIM card and connected to the host PC via USB to enable the ADB interface. The detailed information of the tested devices and their baseband versions are summarized in the Table~\ref{tbl:device_info}.

\begin{table*}[t]
\centering
\caption{5G Smartphones Used in Testing.}
\label{tbl:device_info} 
\vspace{2ex} 
\setlength{\tabcolsep}{6pt}   
\renewcommand{\arraystretch}{1.1}  
\begin{tabular}{@{}l l l l l@{}}
\toprule
\textbf{Vendor} & \textbf{Model} & \textbf{CPU} & \textbf{Baseband Version} & \textbf{Kernel} \\
\midrule
OnePlus Nord CE 2 & IV2201    & Mediatek Dimensity 900  & M\_V3\_P10 & 4.14.186 \\
Redmi K40         & M2012K10C & Mediatek Dimensity 1200 & \makecell[tl]{MOLY.NR15.R3.TC8.PR2\\ .SP.V2.1.P70} & 4.14.186 \\
Vivo X90          & V2241A    & Mediatek Dimensity 9200 & \makecell[tl]{MOLY.NR16.R2.MP1.TC19\\ .PR1.SP.V1.P147} & 5.15.149 \\
Vivo X100         & V2309A    & Mediatek Dimensity 9300 & \makecell[tl]{MOLY.NR16.R2.MP3.TC19\\ .PR1.SP.V1.P124} & 6.1.84 \\
Asus ROG Phone 5s & ASUS\_I005\_1 & Qualcomm Snapdragon 888     & M3.13.24.85-Anakin2 & 5.4.147 \\
OnePlus 13        & PJZ110    & Qualcomm Snapdragon 8 elite & Q\_V1\_P14 & 6.6.30 \\
Pixel 6           & Pixel 6   & Google Tensor              & \makecell[tl]{g5123b-135085-240517-B-\\11857288} & 5.10.198 \\
Pixel 8a          & Pixel 8a  & Google Tensor G3           & \makecell[tl]{g5300o-240308-240517-B-\\11857457} & 5.15.137 \\
\bottomrule
\end{tabular}
\end{table*}

\section{From Evidence to Rule: DSL Synthesis}

\paragraph{Unified DSL induction from specification evidence.}
For each target pair of fields, we assemble a small evidence package that contains the exact ASN.1 block of the IE, the canonical field names, and verbatim sentences from 3GPP texts that co-mention the two fields. A unified inference step reads only this package and decides to return either \texttt{NO\_RULE} or a normalized DSL rule. Admission is strict. The evidence must include clear normative cues such as ``shall'', ``must'', or ``only if'', or an explicit mathematical or tabular relation. The two fields must lie within the same IE instance for this pass. The rule must state the trigger and the effect, stay within the given ASN.1 domains, and normalize units and enumerants when needed. Version or option guards are recorded as preconditions when the text limits scope. Accepted rules are expressed in a small predicate language with two families, \textit{ValueDependency} and \textit{RangeAlignment}, and carry minimal citations to the supporting sentences. Items with only advisory wording, empty or tautological conditions, or missing co-mentions are rejected as \texttt{NO\_RULE}.

\paragraph{Execution and minimal edits.}
Given a decoded message $M$ and a DSL rule $r = (A \Rightarrow C)$, the engine plans a minimal change on the named fields to either satisfy $r$ or deliberately violate it. Candidate edits are checked against the ASN.1 domains so that the re-encoded message $M^{\prime}$ remains well-formed. If the predicate cannot be satisfied or violated within the declared domains for the current instance, the pair is marked infeasible and skipped. Value range and presence probes are produced deterministically from the schema, while intra-IE and inter-IE dependencies use the DSL predicate to guide concrete assignments. Each accepted edit is applied locally, recorded with its field paths and values, re-encoded with \texttt{pycrate}, and delivered through the same execution pipeline.



\section{Prompt for Evidence Bound DSL Induction}
\label{sec:prompt_dsl}

\begin{figure}[th]
    \centering    
    \includegraphics[width=\linewidth]{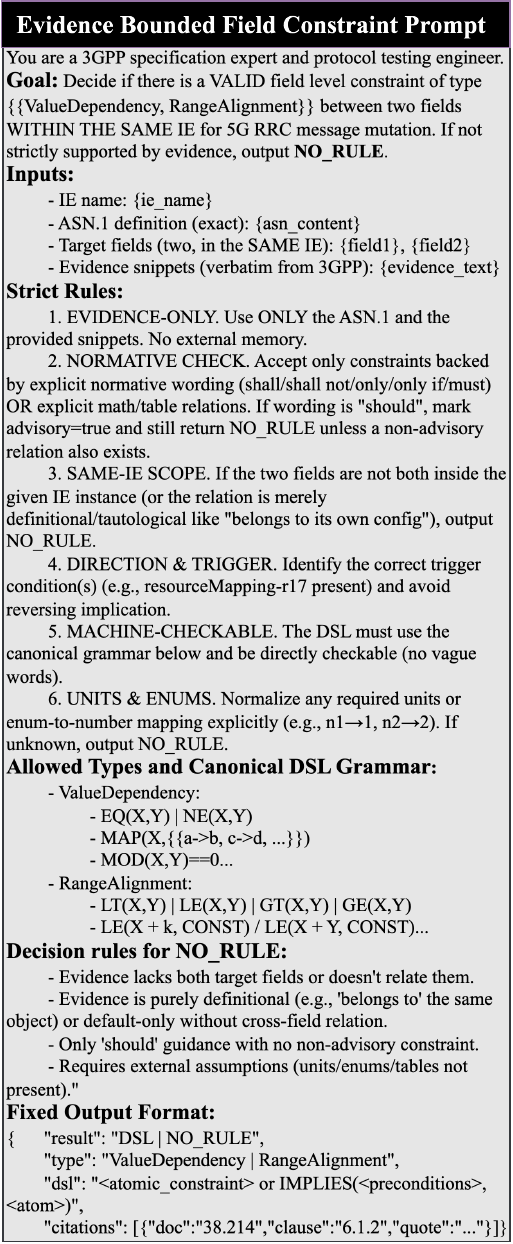}
    \caption{Prompt For intra-IE DSL Rule Induction.}
    \label{fig:prompt_dsl}
\end{figure}

We use an evidence-bound prompt to turn specification text into one executable rule. The prompt takes four inputs: the IE name, the exact ASN.1 block for that IE, the two target fields inside the same IE, and short verbatim snippets from 3GPP that mention the fields. The model must either return \texttt{NO\_RULE} or one normalized DSL clause together with a minimal citation. The prompt enforces six gates that control correctness: evidence only, a normative wording check, same IE scope, explicit direction and trigger, a machine checkable DSL form, and explicit normalization of units and enumerants. If any gate fails, the output must be \texttt{NO\_RULE}. This prevents the use of external knowledge and avoids vague or non-executable text. The allowed grammar covers two families. ValueDependency includes equality and inequality, small maps between enumerants, and simple modular relations. RangeAlignment includes order relations and linear bounds that relate two fields or one field and a constant. The output schema is fixed and compact: a short JSON object with \texttt{result}, \texttt{type}, \texttt{dsl}, and \texttt{citations}. This schema supports programmatic checks and links every rule back to the cited specification sentence. Details are shown in Figure~\ref{fig:prompt_dsl}.

\section{Findings from Specification and Implementation}


\noindent\textbf{Ambiguity in field descriptions.}
We observed frequent ``may only \ldots'' phrasing across TS~38.331 field descriptions.
While such wording is commonly used to express restrictive conditions in 3GPP specifications,
it does not always provide an explicit, machine-checkable trigger for enforcement.
As a result, we treat these descriptions as suggestive rather than definitive evidence,
require ASN.1 satisfiability, and rely on execution feedback before admitting a rule.



\noindent\textbf{Implementation Limits and Unsupported Options Trigger Crashes.} During experiments on open-source UE codebases, \texttt{gdb} backtraces showed that several crashes were terminations triggered by explicit assertions guarding unimplemented or unsupported features. For example, we observed failures at:

\begin{lstlisting}[language=C, numbers=left, xleftmargin=0.5cm, basicstyle=\ttfamily\small, breaklines=true]
AssertFatal(1 == 0,
"Sub-band CQI reporting not yet supported");
\end{lstlisting}

\begin{lstlisting}[language=C, numbers=left, xleftmargin=0.5cm, basicstyle=\ttfamily\small, breaklines=true]
AssertFatal(pucch->n_csi == 0, 
"Multiplexing periodic CSI on PUSCH not supported\n");
\end{lstlisting}

A scan of the OAI public codebase found \textbf{67} ``not implemented'' assertions that intentionally abort when unsupported options are encountered, whereas our over-the-air tests on commercial smartphones did not show comparable assertion-driven terminations. We treat such assertion-induced stops as robustness issues or developer kill switches rather than attacker-reliably-exploitable vulnerabilities.
This distinction is important for open-source stacks that use assertions to mark incomplete functionality, and it helps separate engineering aborts from exploitable faults when interpreting results.

In particular, some failures observed on OAI UE can be attributed to limited robustness or incomplete handling of boundary conditions, rather than specification-level semantic inconsistencies. We therefore treat such cases as implementation artifacts of open-source stacks, and do not rely on them as primary evidence of semantic vulnerabilities in commercial devices.

\section{Schema-Valid Mutation Baseline under OTA Constraints}
\label{sec:baseline}

Recent state-of-the-art protocol fuzzing systems, such as ChatAFL~\cite{meng2024large} and RANsacked~\cite{bennett2024ransacked},  adopt different testing objectives and feedback models compared to \ProjectName{}. ChatAFL leverages large language models to guide seed generation and mutation strategies for improved coverage, while RANsacked focuses on grammar-aware ASN.1 mutation combined with coverage feedback through an instrumented communication channel. In contrast, \ProjectName{} targets \emph{specification-derived semantic constraints} between protocol fields and deliberately generates syntactically valid but semantically inconsistent messages, without attempting to model complex protocol states or optimize coverage.

As a result, a direct apples-to-apples comparison with coverage-guided fuzzers is not practical in our setting. Our evaluation targets a fully black-box over-the-air environment on both open-source and commercial UEs, where reliable coverage feedback is unavailable due to multi-threaded baseband execution and proprietary firmware boundaries. In addition, OTA testing incurs high per-test overheads, including radio reinitialization and device resets, which fundamentally constrain the feasible execution budget.

\noindent\textbf{Schema-Valid Mutation Baseline.} To enable a meaningful comparison under these OTA constraints, we implement a \emph{schema-valid mutation baseline} as a proxy for grammar-aware mutation without semantic reasoning. While RANsacked provides a black-box fuzzing mode, it assumes a persistent execution harness with explicit liveness signaling, which is infeasible in our pre-authentication OTA setting targeting UE baseband behavior. We therefore implement a schema-valid baseline that generates test cases by enumerating field values within ASN.1-defined domains at the individual field level and re-encoding the resulting messages for transmission, without explicitly modeling or intentionally violating cross-field semantic dependencies.

\noindent\textbf{OTA Evaluation Budget.} We compare \ProjectName{} against the schema-valid mutation baseline under realistic OTA execution constraints. Following the OTA testing configuration adopted in \textsc{5Ghoul}, each transmitted test case is executed for a fixed duration of 45 seconds, which serves as a global timeout to determine whether a UE successfully responds or becomes non-responsive. Each test case requires reinitialization of the 5G test environment (including gNB and UE state) to ensure reproducible execution, making OTA testing inherently costly. Under this execution model, a 24-hour OTA budget corresponds to approximately $N \approx 1551$ transmitted test cases.

\noindent\textbf{Results.} Under a unified OTA execution budget, the schema-valid mutation baseline exposes only a limited number of failure cases, whereas \ProjectName{} uncovers substantially more diverse and interpretable crash behaviors. Although this evaluation setting inherently favors high-throughput grammar-based mutation, the results indicate that constraint-guided semantic testing is more effective at exercising meaningful failure paths under realistic OTA conditions. This advantage stems from explicitly targeting specification-level semantic inconsistencies.

\end{document}